\documentclass{aa}
\usepackage{graphicx}
\usepackage{txfonts}
\usepackage{multirow} 
\usepackage[outdir=./figures/]{epstopdf}
\usepackage{diagbox} 
\usepackage{setspace}
\usepackage{float} 
\usepackage[usenames,dvipsnames,svgnames,table]{xcolor} 
\usepackage[colorlinks,allcolors=blue]{hyperref}
\usepackage{booktabs} 
\usepackage{amssymb, amsmath} 
\usepackage{lipsum}  
\usepackage[none]{hyphenat}
\usepackage{ulem} 
\newcommand\omicron{o} 

\newcommand{\co}{\mbox{CO}}
\newcommand{\sio}{\mbox{SiO}}
\newcommand{\so}{\mbox{SO}}
\newcommand{\doce}{\mbox{$^{12}$CO}}

\newcommand{\trece}{\mbox{$^{13}$CO}}

\newcommand{\kms}{\,km\,s$^{-1}$}

\newcommand{\ms}{\mbox{$M_{\odot}$}}

\newcommand{\secp}{\mbox{\rlap{.}$''$}}
\newcommand{\raqr}{R\,Aqr}
\newcommand{\mas}{\,mas}
\newcommand{\mm}{\,mm}
\newcommand{\cm}{\,cm}
\newcommand{\mJy}{\,mJy}
\newcommand{\fig}{Fig.\,}
\newcommand{\figs}{Figs.\,}
\newcommand{\tab}{Table\,}
\newcommand{\tabs}{Tables\,}
\newcommand{\sect}{Sect.\,}
\newcommand{\sects}{Sects.\,}
\newcommand{\mJyBeam}{\,mJy\,beam$^{-1}$}
\newcommand{\JyBeam}{\,Jy\,beam$^{-1}$}
\newcommand{\x}{$\times$\,}
\newcommand{\tdeg}{\textdegree}

\begin{document}

   \title{Continuum and molecular emission from the inner regions of the symbiotic system R\,Aquarii}

   \author{M. Gómez-Garrido\inst{1,2}, V. Bujarrabal\inst{3}, J. Alcolea\inst{1}, A. Castro-Carrizo\inst{4}, J. Mikołajewska\inst{5}, M. Santander-García\inst{1,2}}

   \institute{Observatorio Astron\'omico Nacional (OAN-IGN), Alfonso XII 3, E-28014, Madrid, Spain\\email:{ m.gomezgarrido@oan.es}   
         \and
         Centro de Desarrollos Tecnológicos, Observatorio de Yebes (IGN), 19141 Yebes, Guadalajara, Spain    
        \and
         Observatorio Astronómico Nacional (OAN-IGN), Apartado 112, E-28803 Alcalá de Henares, Spain
        \and 
                Institut de Radioastronomie Millimétrique, 300 rue de la Piscine, 38406 Saint-Martin-d’Hères, France
                \and 
                Nicolaus Copernicus Astronomical Center, Polish Academy of Sciences, Bartycka 18, PL 00-716 Warsaw, Poland
 }

\authorrunning{Gómez-Garrido, M. et al. }

   \date{Accepted ----} 

 
  \abstract
   {Symbiotic systems often include an asymptotic giant branch (AGB) star and a hot compact companion, such as a white dwarf, that are in close interaction. Due to the intense ultraviolet emission from the hot companion, the molecular content of circumstellar envelopes in the symbiotic systems is poor. As a result, the less abundant molecules have not been previously studied in detail in this kind of object.}
   {\raqr{} is the closest and best-studied symbiotic system. Our aim is to study the inner regions of \raqr{} based on ALMA observations of the continuum and line emission.} 
   {We present very sensitive ALMA maps of the continuum emission at 1.3, 0.9, and 0.45\,mm. We also obtain the spatial distribution of the recombination line H30$\alpha$ with a high and moderate angular resolution, and it is compared with the emission of the continuum at 1.3\,mm. High-resolution maps of several molecules are obtained in the three observed ALMA bands. We study the molecular emissions using a simplified model to explain the brightness distributions seen in the central position of our maps.}
   {We find that the low-resolution continuum map at 1.3\,mm shows the emission of the radio photosphere of the AGB star, its surroundings, and the structure of the bipolar jet launched by the companion. The high-resolution continuum map at 1.3\,mm shows the innermost part of the jet, probably revealing the position of the secondary, and suggests mass transfer from the AGB star to the white dwarf. The brightness distribution of H30$\alpha$ is similar but not coincident with the continuum emission, and it basically probes the region where the jet is formed. The brightness distributions of the studied molecular lines show a variety of shapes. The emissions of the abundant molecules, \co{} and \sio{}, are relatively extended since they can survive far from the AGB star in spite of the intense ultraviolet emission from the white dwarf. On the contrary, less abundant molecules only survive in regions close to the AGB star, where shielding is stronger. From our best-fit model for these weak species, we find that the expanding shell is $\sim$10$^{14}$\cm{} in size, which means that these less abundant species are confined to the intra-orbital regions.} 
   {}


   \maketitle

\section{Introduction}\label{sec.introduction}

Symbiotic stellar systems are interacting binaries where the primary component is a cool evolved star, often an asymptotic giant branch (AGB) star, and the secondary is a hot compact object \citep{mikolajewska12}. The material ejected by the AGB falls to the hot companion, resulting in the formation of an accretion disk, which can eventually lead to the ejection of bipolar collimated outflows or jets. These jets interact with the circumstellar envelope of the AGB star, resulting in bipolar structures very similar to those found in many planetary nebulae (PNe). Therefore, studying the distribution of gas and dust in these binary systems can help provide insight into the origin of the axial symmetry in planetary nebulae in general.

\raqr{} is the best studied D-type symbiotic system. It has been observed with high resolution at many wavelengths, and its structure has been studied in detail. The primary component of the stellar system is a Mira-type variable with a pulsation period of 388\,days and a M6.5--8.5 spectral type. Based on very high spatial resolution observations at 0.9\mm{} and near-infrared (NIR) \citep[see][]{bujarrabal18,wittkowski16,ragland08}, we assumed an angular size of $\sim$15\,mas for the primary component. The mass-loss rate of the primary is between 10$^{-6}$ and 10$^{-5}$\,\ms yr$^{-1}$ \citep[see][]{bujarrabal10,bujarrabal21}. The secondary component is a white dwarf (WD). Due to the low luminosity of the WD, it is difficult to directly observe it. Moreover, the information about this component is poor.

\begin{table*}[htpb]
\centering\caption{Spectral configuration for every observed spectral window.}
\begin{tabular}{c c c r c}
\hline \noalign{\smallskip}
\multirow{2}{*}{ALMA Band} & Spectral & Central rest frequency  & Bandwidth & Nominal spectral resolution     \\
 & Window & (GHz) & (MHz) & (MHz/channel) \\
\noalign{\smallskip}
\hline \noalign{\smallskip}

\multirow{4}{*}{6} & 25 & 230.538 & 468.75  & 0.282 \\
& 27 & 231.350 & 1875.00 & 1.129 \\
& 29 & 220.398 & 468.75  & 0.282 \\
& 31 & 216.500 & 1875.00 & 1.129 \\ \midrule
\multirow{4}{*}{7} & 25 & 345.796 & 468.75  & 0.282 \\
& 27 & 343.500 & 1875.00 & 1.129 \\
& 29 & 330.588 & 468.75  & 0.282 \\
& 31 & 331.300 & 1875.00 & 1.129 \\ \midrule
\multirow{4}{*}{9}& 25 & 691.473 & 937.50  & 0.564 \\
& 27 & 693.600 & 1875.00 & 1.129 \\
& 29 & 689.000 & 1875.00 & 1.129 \\
& 31 & 674.800 & 1875.00 & 1.129 \\
\hline
  \end{tabular}
  \label{tab.freqObs}
\end{table*}

In the visible and infrared ranges, the large-scale picture of \raqr{} is dominated by an axisymmetric extended nebula 2\arcmin{} in size with its north side slightly inclined toward the observer \cite[see][]{solf85,schmid17,melnikov18}. The nebula image at visible and IR frequencies shows an equatorial structure elongated almost in the east-west direction. This structure is compatible with a ring-like shape tilted about 20\tdeg{} with respect to the plane of the sky. As a result of the interaction between the jet and the material ejected by the AGB star, two bipolar lobes appear in the direction perpendicular to the equatorial structure. These lobes show point symmetry, suggesting that the jet is precessing with a position angle (PA) ranging between $-$10\tdeg{} and 45\tdeg{}. At the smallest scales in the optical, the images are dominated by the AGB photosphere and the base of the jet launched by the companion; the emission of the atmosphere of the WD is very weak \citep{schmid17}.

Due to the intense UV emission from the WD, most of the molecules are photodissociated in symbiotic systems. Before our work, only a few molecules, such as \co{}, \sio{}, and H$_{2}$O had been detected in some stellar systems, including \raqr, $\omicron$\,Cet, CH\,Cyg, and V627\,Cas \citep[see][]{bujarrabal10,cho10,ramstedt14}. Some symbiotic systems have been mapped using the maser emission of H$_2$O and SiO \citep[][]{yang20,yang21}. Because the formation of maser emission takes place under very specific physical conditions, the maps of SiO or H$_2$O maser lines do not provide an overview of how the molecule-rich gas is distributed in symbiotic systems. The thermal emission of \co{}, which gives more realistic information about the molecular gas distribution, has only been studied in \raqr{} and $\omicron$\,Cet \citep[][]{ramstedt14,ramstedt18,bujarrabal18,bujarrabal21}. In this source, the \co{} emission has been found to be forming an equatorial structure smaller than 2\arcsec{} in size, elongated in the east-west direction, which is very different from what has been found in standard AGB stars. \cite{bujarrabal21} modeled the \co{} emission in \raqr{} based on hydrodynamical calculations, concluding that the molecular gas is distributed in a central component as well as in several spiral arcs along the orbital plane of the system.

The first determinations of the orbital parameters of the \raqr{} system were based on the velocity curve of the primary and on uncertain assumptions on the position of the secondary component \citep[see][]{hollis97,ragland08,gromadzki09}. Based on VLT/SPHERE observations, \cite{schmid17} provided the first true direct measurement of the WD location, measuring a distance in the sky of $\sim$45\mas{} between both components. \cite{bujarrabal18} also determined the position of the secondary with respect to the primary using the very high resolution continuum Band~7 ALMA observations. With these positions and the radial velocity curve provided by the SiO maser lines, the relative orbital parameters have recently been improved \citep{alcolea23}. The derived orbital period is 42.4\,yr, which is slightly shorter than previous estimations \citep[e.g.][]{gromadzki09}. The orbit is inclined 110\tdeg{} with respect to the plane of the sky, and its projection on the plane of the sky shows a PA of 95\tdeg{}, with a semi-major axis of 57$\pm$8\mas. The equatorial structure seen in observations of \raqr{} in the optical, IR, and radio ranges agrees with the predicted geometry of the orbit. Furthermore, similar orbital parameters were used in \cite{bujarrabal21} for the modeling of the CO emission, obtaining successful results.

The distance to \raqr{} is uncertain \citep[see discussion in][]{bujarrabal21}. Since the measurements of parallaxes are probably affected by the orbital movements of the system, the obtained distances using this method are discrepant \citep[e.g.][]{gaiaColl23,min14,andriantsaralaza22}. In this work, we assumed a distance of 265$\pm$20\,pc for our calculations, which is derived from the luminosity-period relation \citep{alcolea23}.   

In this paper we present continuum and line emission maps observed in \raqr{} with ALMA, including molecular emission and radio recombination lines (RRLs). We focus on the very inner region of this symbiotic system, studying the material located at distances comparable to the orbital size.

\begin{table*}[htpb]
\centering\caption{Coordinates of the derived centroids of the continuum maps.}

\begin{tabular}{ l c c r r c}
\toprule\noalign{\smallskip}
\multirow{2}{*}{Band}& \multirow{2}{*}{Date} &   Characteristic angular  & \multicolumn{2}{c}{Obtained centroid position}  & ALMA  \\
 &   & resolution (mas) & \makebox[2.5cm][c]{R.A.} &  \makebox[2.5cm][c]{Dec.}  & Configuration \\
\noalign{\smallskip}
\hline \noalign{\smallskip}

Band~6 - Low resolution & June 2019  & 200  & 23:43:49.5004 & -15:17:04.782 & C43-6 \\ 
Band~6 - High resolution & September 2019  & 25   & .4993 & .769 & C43-10 \\ 
Band~6 - Low + High res. & --  & 40  & .4994 & .765 \\ %
Band 7             &   November 2017      & 30   & .4962 & .720 & C43-8  \\
Band~9                  & August 2019  & 30   & .4997 & .773 & C43-7  \\
\hline
  \end{tabular}
  \tablefoot{The date of the observations and the characteristic angular resolution of every dataset are also indicated.}
  \label{tab.conts.prel}
\end{table*}

\section{Observations}\label{sec.observations}

We present ALMA maps of the continuum and several emission lines in the symbiotic system \raqr{}. We explored the emission of this object in the ALMA Bands~6,~7, and\,~9, corresponding to 1.3, 0.9, and 0.45\,mm wavelengths (ALMA projects 2018.1.00638.S and 2017.1.00363.S). Angular resolutions between 20 and 250\mas{} were attained depending on the observed frequency, array configuration, and  mapping weighting. The observations of Band~6 were carried out in June and September 2019 with high- and low-spatial resolution, and Bands~7 and 9 were observed in November 2017 and August 2019, respectively. 

Using high spectral resolution units of the correlator, we reached a spectral resolution of 0.282\,MHz per channel at 1.3 and 0.9\mm\, and 0.564\,MHz per channel at 0.45\mm, corresponding to velocity resolutions of 0.38\kms{} in Band~6 and 0.25\kms{} in Bands~7 and 9. Moderate spectral resolution units were also connected to study the distribution of the continuum emission. The lines detected in these lower-velocity resolution units were also studied. In \tab\ref{tab.freqObs}, we report the observed central rest frequencies, the bandwidth, and the nominal spectral resolution for each spectral unit.

The calibration of the 2017.1.00363.S data has been described in \cite{bujarrabal18}, and that of 2018.1.00638.S data was very similar, including in the choice of calibrators. For the absolute flux scaling, J0006-0623 was used as a calibrator. In Band~6 we adopted fluxes of $\sim$1.45\,Jy in June 2019 and $\sim$1.32\,Jy in September 2019. In Band~7, we adopted a flux of $\sim$0.9\,Jy in November 2018, and in Band~9, we took a flux of $\sim$4.3\,Jy in August 2019. Flux uncertainties in Bands~6 and 7 are estimated to be between about 5\% and 8\%, while it is $\sim$15\,\% for Band~9. The calibration of the visibilities was performed using the standard ALMA pipeline in the CASA software. Self-calibration of the \textit{uv} data using the continuum emission was performed by means of GILDAS. Exceptionally, the spectral window 31 of Band~6 was self-calibrated using the maser emission of the SiO $v$=1\,$J$=5--4 line, which shows a flux density larger than 6\JyBeam{}. The components used as position reference for self-calibration were very compact and had an S/N above 500, which guarantees an optimal self-calibration. Our final maps show a high dynamic range, thus allowing for the study of the brightness distribution of the material around \raqr{} in great detail.

The position of the AGB star and the WD varies in time because of the proper motions of the system \citep[with somewhat uncertain values of around 30 and $-$30\,\mas/yr in R.A. and Dec. respectively;][]{gaiaColl23,min14}, and the orbital movements of both components. This yields a complex dynamical scenario. To present our data in the same reference framework, we decided to recenter all of our \textit{uv} tables to the centroid of the continuum emission on each ALMA band, which approximately gives the position of the Mira component. As a result, our final maps are expected to be referenced to the AGB position for all observations \citep[see discussion in][]{bujarrabal18,bujarrabal21}, which helps better compare the maps obtained on the different dates. The coordinates of these centroids were derived from a preliminary analysis of the \textit{uv} tables of the continuum data using the routine \textit{uv\_fit} of the GILDAS packages and are given in Table\,\ref{tab.conts.prel}.

In the particular case of the observations at 1.3\,mm, differences around 17 and 13\,mas were found in right ascension (R.A.) and declination (Dec.) of the continuum centroids, respectively, between the low- and high-resolution observations. These differences are comparable to the position accuracy and much smaller than the beam size of the low-resolution data ($\sim$200\mas), so both positions are compatible (we note that in four months, both proper and orbital motions will produce even smaller changes in the position of the primary). Accordingly, we merged both the high- and low-resolution Band~6 data to increase the S/N and recover any possible flux lost in the most extended configuration of the interferometer. For the merged data, we also computed the coordinates of the continuum centroid, which differs by less than 4\mas{} from the high-resolution one, and which is again within the accuracy of the determination of these positions (see \tab\ref{tab.conts.prel}).

\begin{figure}[htpb!]
   \centering{\resizebox{9cm}{!}{
   \includegraphics{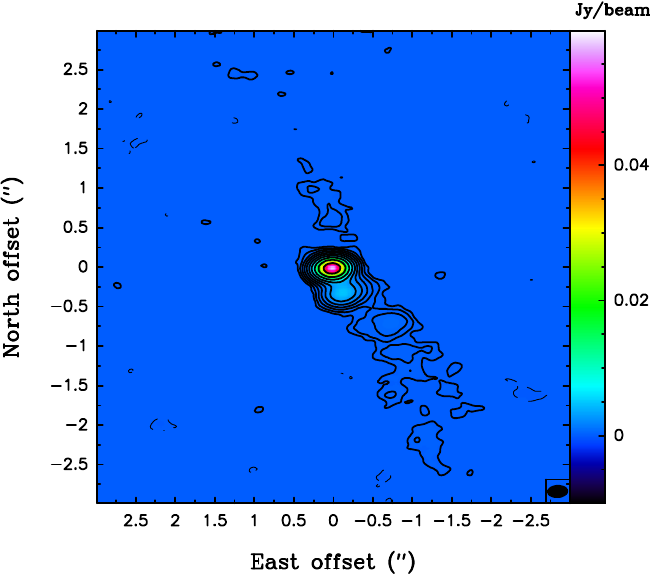}}}
   \caption{Low-resolution 1.3\mm{} continuum map toward \raqr. The contour level spacing is logarithmic with a jump of a factor of two. The first contour is $\pm$0.07\mJyBeam{}. Dashed lines indicate negative levels. The image is centered on the continuum centroid. The beam, with a size of 264\,\x163\mas{} and PA of 95\,º, is shown in the lower-right corner inset.}
    \label{fig.map.conB6Low}
    \end{figure}

\begin{figure}[htpb!]
   \centering{\resizebox{9cm}{!}{
   \includegraphics{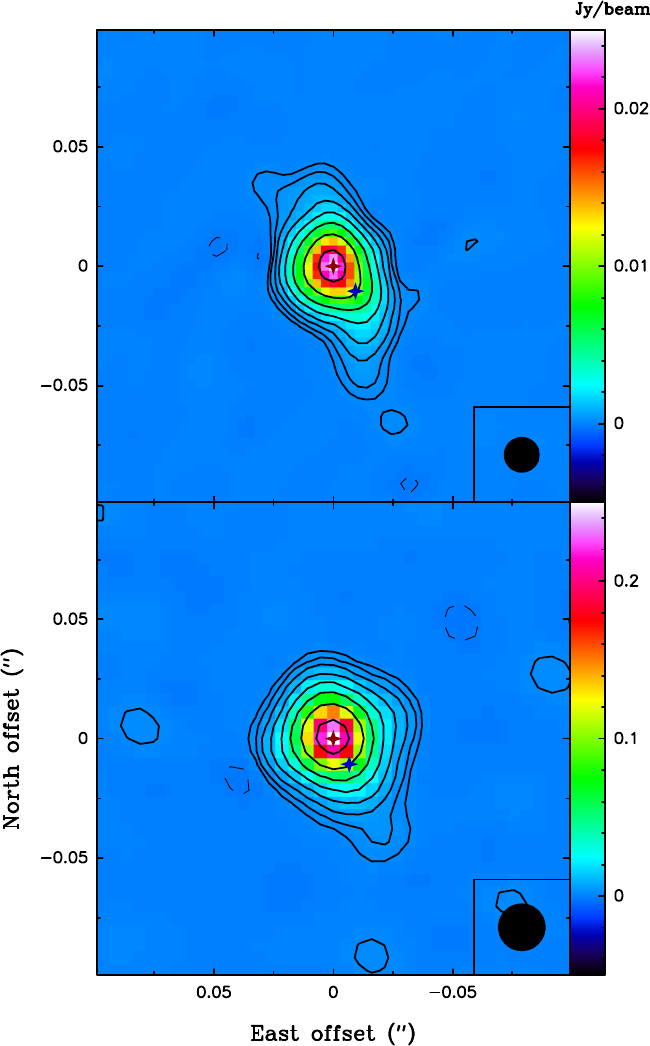}}}
   \caption{High-resolution continuum maps at 1.3 (top panel) and 0.45\mm{} (bottom panel) toward \raqr. The contour level spacing is logarithmic, with a step of a factor of two. The first contours are $\pm$0.15 and $\pm$1.5\mJyBeam{} for the 1.3 and 0.45\mm{} maps, respectively. Dashed lines indicate negative levels. The observations are centered at the continuum centroid and taken as the position of the AGB primary (indicated by a red star). The expected position for the secondary is marked by a blue star. The restoring beam shape is displayed in the lower-right corner inset.} 
    \label{fig.map.conB6B9High}%
    \end{figure}    

The final maps at 1.3 and 0.45\,mm are centered at the International Celestial Reference System (ICRS; J2000) coordinates R.A.: 23:43:49.4994, Dec.: –15:17:04.765 and R.A.: 23:43:49.4997, Dec.: –15:17:04.773, respectively. The ICRS coordinates for the maps at 0.9\,mm are R.A.: 23:43:49.4962, Dec.: –15:17:04.720. The different coordinates obtained in the different datasets can mainly be explained by the proper motions of the system. The orbital motion of the primary and the parallax can also play a relevant role (see \sect\ref{sec.introduction}).

To better study the line emission, the continuum level is subtracted. To subtract the continuum emission, firstly, channels without line emission are selected. The emission in these channels is averaged to obtain the continuum emission in each spectral window. The continuum is subtracted from the \textit{uv} data affecting the visibilities directly. For the frequencies and bandwidth used in this work (see \tab\,\ref{tab.freqObs}), and assuming a spectral index of ~two, differences in the intensity of the continuum emission smaller than 1\,\% are expected for the extrema of the different spectral windows. Therefore, the continuum can be considered constant in each spectral window. We verified that the continuum was properly subtracted, checking many channels without line emission. In these channels, we checked that there is no positive or negative emission after the continuum subtraction. As discussed in \sects\ref{sec.ana.obser} and \ref{sec.models}, the continuum subtraction may affect the brightness distribution observed in the line maps in a complex way. 

Map cleaning was performed in GILDAS using the Hogbom algorithm implemented in MAPPING. We completed a first cleaning using a maximum number of clean components of $\sim$4000. For these first maps, we assumed a stopping criterion of 2.5\,\% of the peak intensity of the dirty image for the maximum amplitude value of the residual map. This yielded preliminary clean maps of the emission lines. Final clean maps were obtained imposing a maximum amplitude of the absolute value of the residual map of $\sigma_{p}$/2, with $\sigma_{p}$ being the noise calculated in the preliminary clean maps. In this case, we increased the maximum number of clean components to 5000. We checked that the final clean maps do not show overcleaned regions. Noise levels of $\sim$0.1, 2, and 10\,mJy were found in the clean maps on Bands\,6, 7, and 9, respectively. Thanks to the good quality of the data, we worked to reach our best angular resolution using robust weighting. Beam sizes of about 25$\times$18, 35$\times$25, and 33$\times$28\mas{} and PAs of around 129, 56, and 83\,deg were achieved for the ALMA Bands~6,~7, and\,~9. However, the synthesis imaging process for the spectral windows 27 and 29 in Band~6 are shown with natural weighting due to the limited S/N in these datasets. As a result, the maps of the $^{12}$CO $v$=0 $J$=2--1, $^{13}$CO $v$=0 $J$=2--1, and AlOH $J$=7--6 lines were computed with a synthesized beam size of 101\,\x72 and 115\,\x80\mas{}, respectively. 

Tables\,\ref{tab.lines.b6}, \ref{tab.lines.b7}, and \ref{tab.lines.b9} collect the frequency and energy of the upper level for each detected transition, ordered by increasing frequency. The Jet Propulsion Laboratory catalog and the Cologne Database for Molecular Spectroscopy were used for line identification. We considered simple molecules, many of them were previously detected in O-rich stars \citep[see e.g.][]{kaminski13,saberi22}, with transitions close to our detections. In some cases the identifications are uncertain or tentative; this is indicated with a (?) note after the line name. In several cases, including some lines with S/N high enough for mapping, we have not found reasonable candidates to attribute the emission. In these cases, the name of the line is coded as "U" followed by the derived rest frequency in megahertz (see, e.g., U343294 and U675616). For the assignation of these frequencies, we have always assumed a systemic velocity of --26\kms{}. Almost all detected lines are due to molecular species. Only two recombination lines were detected, H30$\alpha$ and probably C30$\alpha$, but only H30$\alpha$ is strong enough to be mapped.

Taking into account the intensity of the emissions and the velocity range covered by the detected lines, the maps were derived using a spectral resolution of 1.6\kms{} for all mapped lines except for the recombination line H30$\alpha$. Since the emission of this line covers a wider velocity range (around 110\kms) and its intensity is relatively weak, we resampled the \textit{uv} tables to obtain a spectral resolution of 4.8\kms{}, increasing the S/N.      

We also studied the continuum emission of \raqr. Because the high- and low-resolution continuum maps at 0.9\mm{} were already studied by \cite{bujarrabal18} in detail, we focused on the analysis of the observations at 1.3 and 0.45\mm. 

In order to make our study of the innermost regions of the continuum emissions in \raqr{} easier, we analyzed the high-resolution data imposing a circular restoring beam. We were able to achieve restoring beams of 15 and 20\,mas for the continuum maps at 1.3 and 0.45\,mm, respectively. Because the continuum emission is observed with high S/N, the obtained results were probed to be reliable \citep[see discussion in][]{bujarrabal18}. In addition, we derived low-resolution maps using natural weighting to study the extended continuum emission of \raqr{}. This yielded a resolution of 264\,\x163\mas{}. This extended continuum emission is not detected at 0.45\,mm, where only the compact emission of Mira's radio-photosphere and the innermost jet components are present.

We expected a moderate loss of flux due to over resolution in our maps. Even in the molecules that show a very extended structure, such as \co{}, we only expected a lost flux percentage smaller than $\sim$30\% \citep{bujarrabal21}. This value is expected to be smaller for the compact emission detected in other molecules. The flux lost in the continuum maps at 1.3\mm{} is difficult to evaluate, as very extended and weak emission seems to appear in \fig\ref{fig.map.conB6Low}.

\begin{table}
\caption{Band~6 detected lines: Frequency and synthetized beam. The question mark after the line name indicates uncertain identification.}
\begin{small}
\begin{tabular}{ l l c }
\hline \noalign{\smallskip}
\multirow{2}{*}{Line} & Frequency & E$_u$  \\
     &  (MHz)    &  (K)    \\
\noalign{\smallskip}
\hline \noalign{\smallskip}
SiO $v$=1 $J$=5--4                  &  215596.018 & 1800  \\
SiO $v$=0 $J$=5--4                  &  217104.919 & 31.3  \\
AlOH $J$=7--6 (?)                               &  220330.583 & 42.3  \\
$^{13}$CO $v$=0 $J$=2--1            &  220398.684 & 15.9  \\
CO $v$=0 $J$=2--1                   &  230538.000 & 16.6  \\ 
H30$\alpha$                             &  231900.928 & --    \\
C30$\alpha$ (?)                                         &  232016.636 & --    \\

  \end{tabular}
  \label{tab.lines.b6}
  \end{small}
\end{table}

\begin{table}
\caption{Band~7 detected lines: Frequency and synthetized beam. The question mark after the line name indicates uncertain identification.}
\begin{small}

\begin{tabular}{ l l c }
\hline \noalign{\smallskip}
\multirow{2}{*}{Line} & Frequency & E$_u$  \\
     &  (MHz)    &  (K)    \\

\noalign{\smallskip}
\hline \noalign{\smallskip}
SiO $v$=7 $J$=8--7                                                              &  330477.533   & 12097   \\
$^{13}$CO $v$=0 $J$=3--2                                                        &  330587.965   & 31.7    \\
H$_2$O $v_2$=2 $J_{K_{a},K_{c}}$=$3_{2,1}$-- $4_{1,4}$          &  331123.730   & 4881    \\
TiO$_2$ $J_{K_{a},K_{c}}$=27$_{8,20}$--27$_{7,21}$              &  331211.764   & 346.4   \\
TiO$_2$ $J_{K_{a},K_{c}}$=26$_{8,18}$--26$_{7,19}$              &  331240.149   & 326.6   \\
TiO$_2$ $J_{K_{a},K_{c}}$=39$_{8,32}$--39$_{7,33}$              &  331381.839   & 643.6   \\
$^{30}$SiO $v$=3 $J$=8--7                                                       &  331954.539   & 5264    \\
Si$^{17}$O $v$=1 $J$=8--7                                                       &  332021.994   & 1807    \\
$^{12}$CO $v$=1 $J$=3--2                                                &  342647.636   & 3117    \\
$^{34}$SO $J_{K_{a},K_{c}}$=55$_{6,50}$--56$_{3,53}$ (?)    &  342756.396   & 1509    \\
$^{29}$SiO $v$=0 $J$=8--7                                                       &  342979.109   & 74.1    \\
TiO$_2$ $J_{K_{a},K_{c}}$=11$_{8,4}$--11$_{7,5}$                &  343004.056   & 118.3   \\
TiO$_2$ $J_{K_{a},K_{c}}$=23$_{3,21}$--22$_{2,20}$              &  343071.530   & 208.3   \\
U343294                                                                                                 &  343294       & --      \\
U343416                                                                                                 &  343416       & --      \\
U343797                                                                                                 &  343797       & --      \\
TiO $v$=2 $J_{K_{a},K_{c}}$=11$_{2,211}$--10$_{2,210}$ (?)      &  344224.605   & 2962   \\
SO $^3\Sigma$ $v$=1 $N_J$=9$_8$--8$_7$                                  &  343829.350   & 78.2    \\
SO $v$=0 $J$=8--7                                                                               &  344310.612   & 87.5    \\
$^{12}$CO $v$=0 $J$=3--2                                                        &  345795.990   & 33.2    \\

  \end{tabular}
  \label{tab.lines.b7}
  \end{small} 
\end{table}

\begin{table}
\caption{Band~9 detected lines: Frequency and synthetized beam. The question mark after the line name indicates uncertain identification.}
\begin{small}
\begin{tabular}{ l l c }
\hline \noalign{\smallskip}
\multirow{2}{*}{Line} & Frequency & E$_u$  \\
     &  (MHz)    &  (K)    \\
\noalign{\smallskip}
\hline \noalign{\smallskip}
C$^{17}$O $J$=6--5                                                                              &    674009.344   & 113.2  \\
U674269                                                                                                 &    674269               & --     \\
TiO $v$=1 $J_{K_{a},K_{c}}$=20$_{2,121}$--19$_{2,120}$ (?)              &    674381.612           & 2078   \\
C$^{34}$S    $J$=14--13 (?)                                                             &    674473.618           & 179.8  \\
U674517                                                                                                                 &    674517               & --     \\
$^{34}$SO $N_J$=16$_{16}$-15$_{15}$ (?)                                         &    674759.674           & 288.5  \\
$^{49}$TiO $J$=20--19 (?)                                                                       &    674801.264           & 640.0  \\ 
SiO $v$=4 $J$=16--15                                                                            &    675006.767           & 7250   \\
U675172                                                                                                                 &    675172                       & --     \\   
$^{34}$SO $N_J$=17$_{16}$-16$_{15}$                                                     &    675312.868           & 280.2  \\
$^{47}$TiO  $J_{K_{a},K_{c}}$=22$_{2,124}$--21$_{2,23}$ (?)     &    675347.984         & 494.1  \\ 
U675616                                                                                                                 &    675616               & --     \\
SO $^3\Sigma$ $v$=0 $N_J$=16$_{16}$--15$_{15}$                          &    688204.630           & 294.0  \\
AlO $N$=18--17                                                                                                   &    688327.284         & 314.0  \\
SO $^3\Sigma$ $v$=0  $N_J$=17$_{16}$--16$_{15}$                                 &    688735.700           & 285.7  \\
U689075                                                                                                 &    689075               & --     \\
SiO $v$=1 $J$=16--15                                                                                    &    689465.444           & 2050   \\
TiO$_2$  $J_{K_{a},K_{c}}$=17$_{10,8}$--16$_{9,7}$                      &    691321.308           & 221.1  \\
AlF $J_F$=21$_{41/2}$-20$_{43/2}$                                                       &    691349.228           & 365.3  \\
CO $v$=0 $J$=6--5                                                                                               &    691473.076           & 116.2  \\
H$_2$O $v_3$=1 $J_{K_{a},K_{c}}$=5$_{7,2}$--6$_{6,1}$                   &    691548.656           & 6881   \\   
SiO $v$=9 $J$=17--16                                                                                    &    691612.136           & 15609  \\
U692727                                                                                                         &    692727       & --     \\
$^{13}$CS $J$=15--14 (?)                                                                        &    693233.747           & 266.2  \\
SiO $v$=0 $J$=16--15                                                                            &    694294.114           & 283.3  \\

  \end{tabular}
  
  \label{tab.lines.b9}
  \end{small}
\end{table}


 \begin{figure*}[htpb]
  \centering{\resizebox{18.2cm}{!}{
  \includegraphics{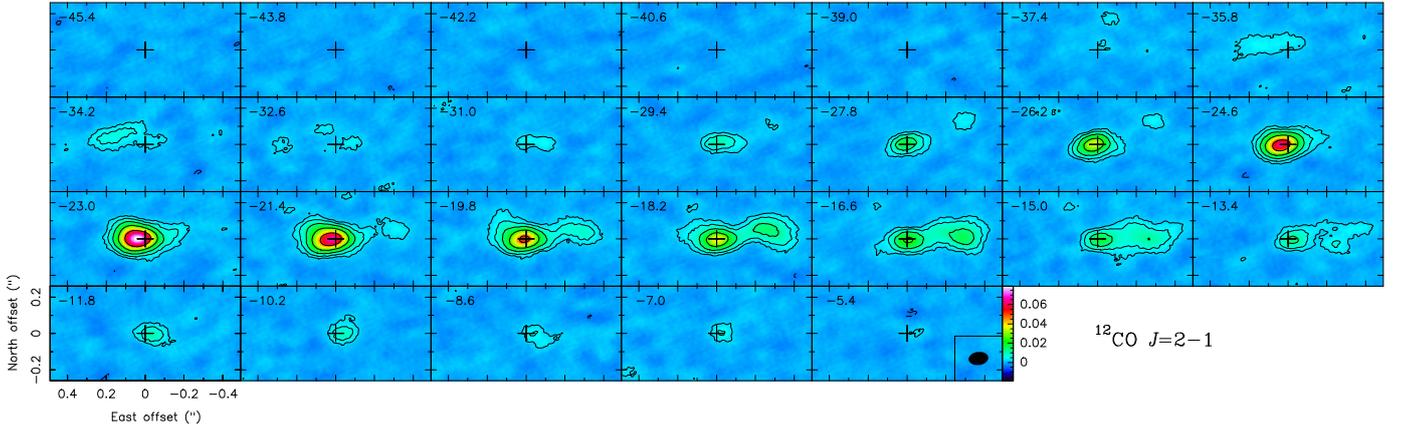}}}
  \caption{Velocity channel maps of $^{12}$CO $J$=2--1 in R Aqr. The LSR velocities are indicated in the upper-left corners. The brightness units are \JyBeam. The contours are logarithmic, with a step of a factor of two and a first level of $\pm$3\mJyBeam. The dashed contours represent negative values. The HPBW is shown in the inset of the last panel.} 
    \label{fig.map.b6_12co21}
\end{figure*}

\begin{figure*}[htpb]
  \centering{\resizebox{18.2cm}{!}{
  \includegraphics{figures/mapLines/b6_sio_v0_j5_4.eps}}}
  \caption{Same as \fig\ref{fig.map.b6_12co21} but for $^{28}$SiO $v$=0 $J$=5--4 using a first level of $\pm$4\mJyBeam.}
    \label{fig.map.b6_sio_v0_j5_4}
\end{figure*} 

\begin{figure*}[htpb]
  \centering{\resizebox{18.2cm}{!}{
  \includegraphics{figures/mapLines/b6_sio_v1_j5_4.eps}}}
  \caption{Same as \fig\ref{fig.map.b6_12co21} but for $^{28}$SiO $v$=1 $J$=5--4 using a first level of $\pm$40\mJyBeam.}
    \label{fig.map.b6_sio_v1_j5_4}
\end{figure*}



 \begin{figure*}[htpb]
  \centering{\resizebox{18.2cm}{!}{
  \includegraphics{figures/mapLines/b7_12co_j3_2.eps}}}
  \caption{Same as \fig\ref{fig.map.b6_12co21} but for $^{12}$CO $J$=3--2 using a first level of $\pm$3\mJyBeam.}
    \label{fig.map.b7_12co_j3_2}
\end{figure*} 

\begin{figure*}[htpb]
  \centering{\resizebox{18.2cm}{!}{
  \includegraphics{figures/mapLines/b7_29sio_v0_j8_7.eps}}}
  \caption{Same as \fig\ref{fig.map.b6_12co21} but for $^{29}$SiO $v$=0 $J$=8--7 using a first level of $\pm$4\mJyBeam.}
    \label{fig.map.b7_29sio_v0_j8_7}
\end{figure*}



\begin{figure*}[htpb]
  \centering{\resizebox{18.2cm}{!}{
  \includegraphics{figures/mapLines/b9_CO_6_5.eps}}}
  \caption{Same as \fig\ref{fig.map.b6_12co21} but for $^{12}$CO $v$=0 $J$=6--5 using a first level of $\pm$2\mJyBeam.}
    \label{fig.map.b9_co_j6_5}
\end{figure*}

\begin{figure*}[htpb]
  \centering{\resizebox{18.2cm}{!}{
  \includegraphics{figures/mapLines/b9_SiOv0_16_15.eps}}}
  \caption{Same as \fig\ref{fig.map.b6_12co21} but for $^{28}$SiO $v$=0 $J$=16--15 using a first level of $\pm$20\mJyBeam. }
    \label{fig.map.b9_sio_v0_j16_15}
\end{figure*}

\begin{figure*}[htpb]
  \centering{\resizebox{18.2cm}{!}{
  \includegraphics{figures/mapLines/b9_SiOv1_16_15.eps}}}
  \caption{Same as \fig\ref{fig.map.b6_12co21} but for $^{28}$SiO $v$=1 $J$=16--15 using a first level of $\pm$20\mJyBeam.}
    \label{fig.map.b9_sio_v1_j16_15}
\end{figure*}

\begin{figure*}[htpb]
  \centering{\resizebox{18.2cm}{!}{
  \includegraphics{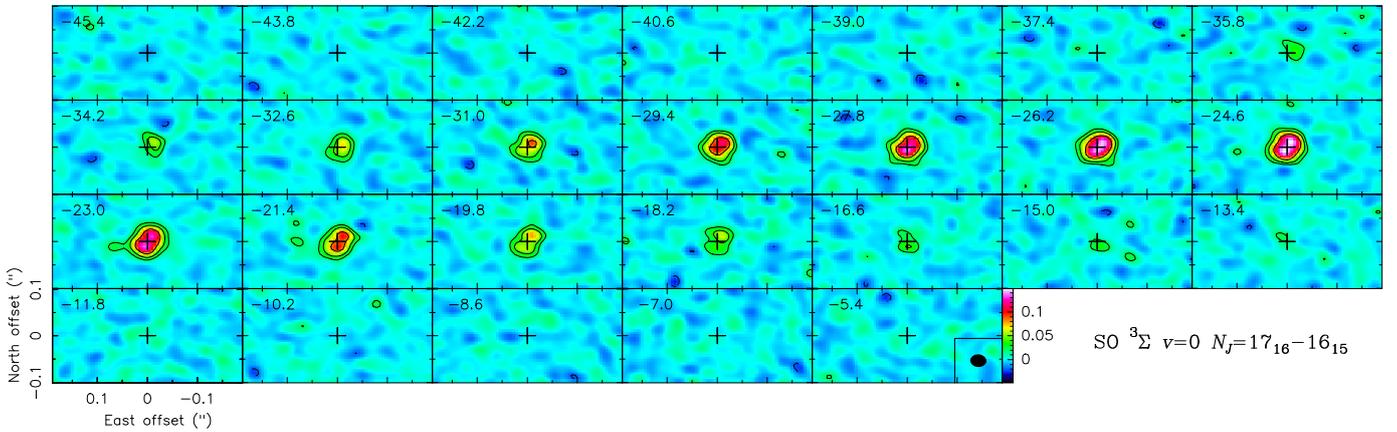}}}
  \caption{Same as \fig\ref{fig.map.b6_12co21} but for SO $^3\Sigma$ $v$=0 $N_J$=17$_{16}$--16$_{15}$ using a first level of $\pm$40\mJyBeam. }
    \label{fig.map.b9_so_v0_17_16}
\end{figure*}

\begin{figure*}[htpb]
  \centering{\resizebox{18.2cm}{!}{
  \includegraphics{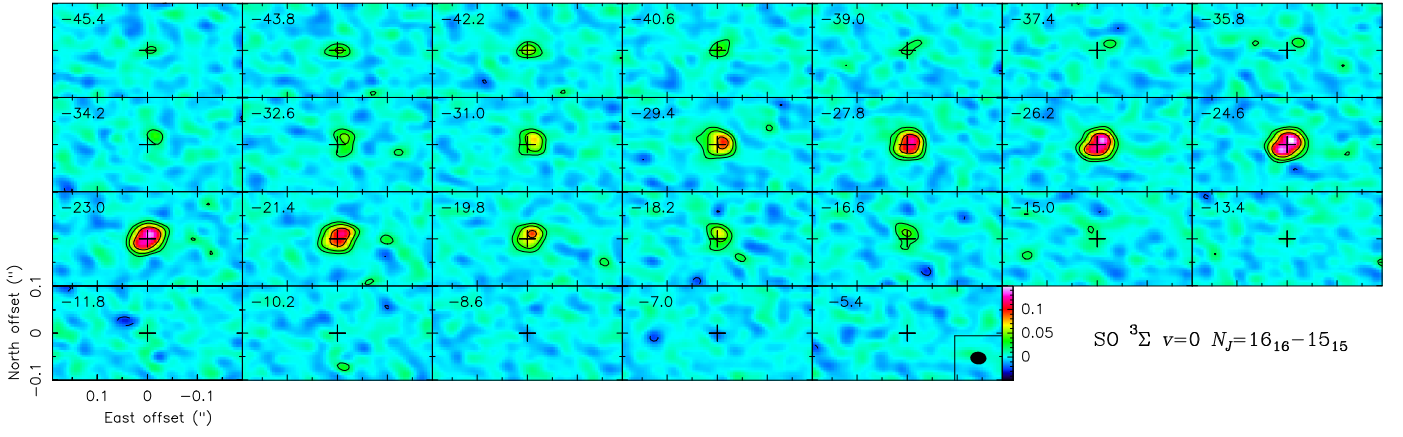}}}
  \caption{Same as \fig\ref{fig.map.b6_12co21} but for SO $^3\Sigma$ $v$=0 $N_J$=16$_{16}$--15$_{15}$ using a first level of $\pm$20\mJyBeam.}
    \label{fig.map.b9_so_v0_j1616_1515}
\end{figure*}


 \begin{figure*}[htpb]
  \centering{\resizebox{18.2cm}{!}{
  \includegraphics{figures/mapLines/b6_h30a.eps}}}
  \caption{ALMA maps per velocity channel of H30$\alpha$ emission in R Aqr. The LSR velocities are indicated in the upper-left corners. The brightness units are \JyBeam. The contours are logarithmic, with a step of a factor of two and a first level of $\pm$1\mJyBeam. The dashed contours represent negative values. The HPBW is shown in the inset in the last panel.}
    \label{fig.map.b6_h30a}
    \end{figure*} 

 \begin{figure*}[htpb]
  \centering{\resizebox{18.2cm}{!}{
  \includegraphics{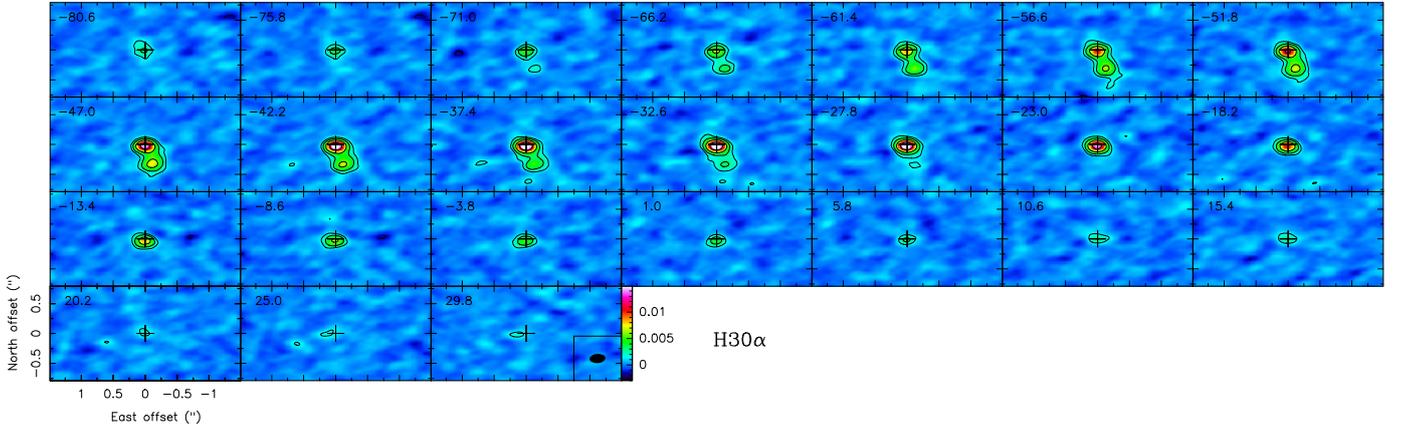}}}
  \caption{Same as \fig\ref{fig.map.b6_h30a} but for low-resolution observations and using a first level of $\pm$1.5\mJyBeam. We note a much larger area is shown in comparison to the previous figure.}
  
    \label{fig.map.b6_h30a_lowres}
    \end{figure*}

\section{Observational results}\label{sec.results}

\subsection{Continuum emission}\label{sec.results.cont}

\fig\ref{fig.map.conB6Low} shows the extended distribution of the continuum emission at 1.3\mm{}. Maps with a higher angular resolution of the continuum emission are in \fig\ref{fig.map.conB6B9High}. We note that the shown area in \figs\ref{fig.map.conB6Low} and \ref{fig.map.conB6B9High} is of 5\arcsec\,\x5\arcsec and 0\secp2\,\x0\secp2, respectively.  

As mentioned before, we assumed that the centroid of the compact continuum emission is also the location of the Mira component (see Sec.\,\ref{sec.observations}). We indicate this position with a red star symbol in \fig\ref{fig.map.conB6B9High}. The predicted relative position of the WD \citep[from orbital parameter by][]{alcolea23} with respect to the AGB is also marked, but with a blue star symbol.

\subsection{Line emission}\label{sec.results.lines}
The detected lines are specified in Tables\,\ref{tab.lines.b6}, \ref{tab.lines.b7}, and \ref{tab.lines.b9}, where the rest frequencies of the observed transitions and the energy of the upper level are indicated. Only lines with a high-enough S/N are mapped. 

Lines observed with a good S/N allowed us to study the spatial distribution of the different velocity components. The shapes detected in these lines, their extent in the R.A. and Dec. directions, and peak brightness are in \tabs\ref{tab.data.b6}, \ref{tab.data.b7}, and \ref{tab.data.b9}.
Some detected molecules show an S/N too low to be studied in maps per velocity channel. These transitions are marked as "low S/N" in \tabs\ref{tab.data.b6}, \ref{tab.data.b7}, and \ref{tab.data.b9}. In these cases, we only report their detections and their measured total intensities.

We present our results using maps per velocity channel. In this way, we could study the different components of the gas at different velocities. In these maps, the first contour is approximately three times the noise level. Maps of the well-resolved lines are presented in this section (see \figs\ref{fig.map.b6_12co21}\,--\,\ref{fig.map.b6_h30a_lowres}). Maps of the lines with a barely resolved spatial distribution and those with a low quality are in Appendix\,\ref{app.maps}. Maps of molecular lines in Band\,6 are shown in \figs\,\ref{fig.map.b6_12co21}, \ref{fig.map.b6_sio_v0_j5_4}, \ref{fig.map.b6_sio_v1_j5_4}, and \ref{fig.map.b6_13co21}. We show maps of lines in Band\,7 (0.9\mm) in \figs\,\ref{fig.map.b7_12co_j3_2}, \ref{fig.map.b7_29sio_v0_j8_7}, and in \figs\,\ref{fig.map.b7_13co_j3_2} to \ref{fig.map.b7_so_v0_j8_7}. Finally, lines mapped in Band\,9 (0.45\mm) are shown in \figs\,\ref{fig.map.b9_co_j6_5} to \ref{fig.map.b9_so_v0_j1616_1515} and in Figs.\,\ref{fig.map.b9_TiO_v1_20_2_121_19_2_120} to \ref{fig.map.b9_C13S}. We show in \figs\ref{fig.map.b6_h30a} and \ref{fig.map.b6_h30a_lowres} maps of H30$\alpha$ with high and low angular resolution; the low-resolution map was obtained by natural weighting and is presented to show the large-scale structure of atomic gas. The maps of the CO transitions $v$=0 $J$=2--1, $J$=3--2, and $J$=6--5, which were shown in \cite{bujarrabal21}, are also reproduced in this paper due to the relevance of this molecule in the study of circumstellar environments and also for comparison purposes. To compare spatial distributions from the different mapped lines, in most plots we have used a map size  of 0\secp4 in R.A. by 0\secp2 in Dec., except for the most extended emissions, for which we have used a map size of 1\secp0\,\x0\secp5.

Due to the high number of lines detected in Band\,9, several lines are often relatively close in frequency. For example, the emission of the TiO$_2$ $J_{K_a,K_c}$=17$_{10,8}$--16$_{9,7}$ and AlF $J_F$=21$_{41/2}$-20$_{43/2}$ lines overlap each other. The emission of TiO$_2$ $J_{K_a,K_c}$=17$_{10,8}$--16$_{9,7}$ can be seen in \fig\,\ref{fig.map.b9_TiO2_17_10_8_16_9_7} in the velocity range characteristic of \raqr{}, between $-$31.0 and $-$18.2\kms{} LSR, while AlF $J_F$=21$_{41/2}$-20$_{43/2}$ appears between $-$40.6 and $-$32.6\kms{}. In Fig.\,\ref{fig.map.b9_AlF_21_20}, we found the emission of AlF $J_F$=21$_{41/2}$-20$_{43/2}$ between $-$29.4 and $-$18.2\kms and the TiO$_2$ $J_{K_a,K_c}$=17$_{10,8}$--16$_{9,7}$ line in the panels labeled between $-$16.6 and $-$5.4\kms. Similar pairs of lines are H$_2$O $v_3$=1 $J_{K_{a},K_{c}}$=5$_{7,2}$--6$_{6,1}$ and SiO $v$=9 $J$=17--16 , and CO $v$=0 $J$=6--5 and H$_2$O $v_3$=1 $J_{K_{a},K_{c}}$=5$_{7,2}$--6$_{6,1}$. In all these cases, only the central velocity channels show the emission from the line indicated in the corresponding figure.

We note that \sio{} maser emission is detected in the rotational transitions $v$=0,1 $J$=5--4. The maps of these lines show unexpectedly extended structures between --26.2 and --19.8\kms{} and --32.6 and --26.2\kms{} for the $v$=0 and 1 transitions, respectively, (see \figs\ref{fig.map.b6_sio_v0_j5_4} and \ref{fig.map.b6_sio_v1_j5_4}). After a detailed self-calibration process using both the continuum and the maser emission, the relatively compact structures were improved in all velocity channels. However, the extended emissions, seen in the velocity ranges mentioned before, remain very similar. We therefore conclude that the "unexpected" extended emission is real and probably has thermal origin.

\begin{table*}
\centering\caption{Band~6 detected lines: shape, extent, and brightness. }
\begin{tabular}{ l c c r r l}
\hline \noalign{\smallskip}
\multirow{2}{*}{Line} &  \multirow{2}{*}{Shape} & Extent & \multicolumn{2}{c}{--- Peak brightness ---} & \multirow{2}{*}{Comments} \\
     &      & (mas)      & (K) &  (\mJyBeam{})    &    \\
\noalign{\smallskip}
\hline \noalign{\smallskip}
SiO $v$=1 $J$=5--4   & extended & 200$\times$100 & 30200 & 520 & maser    \\
SiO $v$=0 $J$=5--4   & extended+ring & 200$\times$100  & 380000 & 6200 & maser  \\
AlOH $J$=7--6   &  -- &  & 10 & 3 & low S/N    \\
$^{13}$CO $v$=0 $J$=2--1   &  extended+ring & 300$\times$200 & 40 & 15 &  \\
CO $v$=0 $J$=2--1  &  elongated east-west & 800$\times$300 & 250 & 80 &  \\ 
H30$\alpha$        &  extended & 40$\times$80  & 320 & 5.5 &  \\
C30$\alpha$       &  -- & & 70 & 1.5 & low S/N  \\
  \end{tabular}
  \label{tab.data.b6}
\end{table*}

\begin{table*}
\centering\caption{Band~7 detected lines: shape, extent, and brightness.}
\begin{tabular}{ l c c r r l }
\hline \noalign{\smallskip}
\multirow{2}{*}{Line} &  \multirow{2}{*}{Shape} & Extent & \multicolumn{2}{c}{--- Peak brightness ---} & \multirow{2}{*}{Comments} \\
     & & (mas)     &  (K) & (\mJyBeam{})     &    \\
\noalign{\smallskip}
\hline \noalign{\smallskip}

SiO $v$=7 $J$=8--7   & -- &     & 80 & 6 & low S/N \\                                         
$^{13}$CO $v$=0 $J$=3--2  & extended+ring & 180$\times$100 & 260 & 20 & \\                    
H$_2$O $v_2$=2 $J_{K_{a},K_{c}}$=$3_{2,1}$-- $4_{1,4}$ & compact &  80$\times$70 & 290 & 22 & \\
TiO$_2$ $J_{K_{a},K_{c}}$=27$_{8,20}$--27$_{7,21}$  &  -- &   & 40 & 4 & low S/N  \\          
TiO$_2$ $J_{K_{a},K_{c}}$=26$_{8,18}$--26$_{7,19}$  & -- &  & 40 & 4 & low S/N  \\            
TiO$_2$ $J_{K_{a},K_{c}}$=39$_{8,32}$--39$_{7,33}$  & -- &  & 30 & 3 & low S/N \\             
$^{30}$SiO $v$=3 $J$=8--7  & compact & 70$\times$80 & 200 & 15 &  \\                          
Si$^{17}$O $v$=1 $J$=8--7  & -- &  & 130 & 10 & low S/N \\ 
$^{12}$CO $v$=1 $J$=3--2  & compact & 200$\times$100 & 230 & 18 & \\                                 
$^{34}$SO $J_{K_{a},K_{c}}$=55$_{6,50}$--56$_{3,53}$  & -- &   & 70 & 6 & low S/N\\             
$^{29}$SiO $v$=0 $J$=8--7  & extended+ring & 450$\times$100 & 400 & 30 & absorption \\        
TiO$_2$ $J_{K_{a},K_{c}}$=11$_{8,4}$--11$_{7,5}$    & -- &  & 60 & 5 & low S/N  \\            
TiO$_2$ $J_{K_{a},K_{c}}$=23$_{3,21}$--22$_{2,20}$   & -- &  & 60 & 5 & low S/N  \\           
U343294      & -- &  & 60 & 5 & low S/N  \\                                                
U343416      & -- &  & 60 & 5 & low S/N  \\                                                
U343797     & -- &  & 60 & 5 & low S/N  \\                                                
TiO $v$=2 $J_{K_{a},K_{c}}$=11$_{2,211}$--10$_{2,210}$   & -- &  & 60 & 5 & low S/N  \\   
SO $^3\Sigma$ $v$=1 $N_J$=9$_8$--8$_7$ & compact & 80$\times$70 & 220 & 17 & \\                
SO $v$=0 $J$=8--7   & extended+peanut  & 90$\times$80 & 330 & 25 & \\                         
$^{12}$CO $v$=0 $J$=3--2  & elongated east-west & 800$\times$200 & 400 & 55 & \\

  \end{tabular}
  \label{tab.data.b7}
\end{table*}

\begin{table*}
\centering\caption{Band~9 detected lines: shape, extent, and brightness.}
\begin{tabular}{ l c c r r l }
\hline \noalign{\smallskip}
\multirow{2}{*}{Line} &  \multirow{2}{*}{Shape} & Extent & \multicolumn{2}{c}{--- Peak brightness ---} & \multirow{2}{*}{Comments} \\
     & & (mas)     &  (K) & (\mJyBeam{})     &    \\
\noalign{\smallskip}
\hline \noalign{\smallskip}

C$^{17}$O $J$=6--5  &  -- &    & 90 & 30 & low S/N \\                                             
U674269    & -- &    & 110 & 40 & low S/N \\                                               
TiO $v$=1 $J_{K_{a},K_{c}}$=20$_{2,121}$--19$_{2,120}$     & compact & 70$\times$90 & 280 & 100\\       
C$^{34}$S $J$=14--13    & -- & &  90  & 30 & low S/N  \\                                                      
U674517  & -- &    & 70 & 20 & low S/N \\ 
$^{34}$SO $N_J$=16$_{16}$-15$_{15}$ & --& 90$\times$80 & 110 & 35  & low S/N \\                      
$^{49}$TiO $J$=20--19    &  compact &  70$\times$70  & 240 & 80 & \\                         
SiO $v$=4 $J$=16--15    & extended+peanut & 100$\times$90 & 550 & 190 & absorption \\     
U675172    & -- & & 100 & 30 & low S/N \\                   
$^{34}$SO $N_J$=17$_{16}$-16$_{15}$   & -- &  & 130 & 50 & low S/N  \\           
$^{47}$TiO $J_{K_{a},K_{c}}$=22$_{2,124}$--21$_{2,23}$  & -- &   & 160 & 60 & low S/N \\
U675616    & extended+peanut & 90$\times$80 & 240 & 80 &  \\  
SO $^3\Sigma$ $v$=0 $N_J$=16$_{16}$--15$_{15}$   & extended+peanut & 100$\times$100 & 450 & 150 \\        
AlO $N$=18--17   & compact &  70$\times$40 & 300 & 110 & \\        
SO $^3\Sigma$ $v$=0 $N_J$=17$_{16}$--16$_{15}$ & extended+peanut & 100$\times$100 & 450 & 150 &  \\       
U689075 & compact &  60$\times$80  & 220 & 90 &  \\                              
SiO $v$=1 $J$=16--15 & extended+ring & 220$\times$110 & 550 & 180 &  \\                   
TiO$_2$ $J_{K_{a},K_{c}}$=17$_{10,8}$--16$_{9,7}$  & compact & 70$\times$70  & 320 & 110 &      \\    
AlF $J_{F}$=21$_{41/2}$-20$_{43/2}$  & compact &  70$\times$60  & 240  &  80 &                               \\
CO $v$=0 $J$=6--5  & extended+ring & 800$\times$200 & 590 & 200 & \\                        
H$_2$O $v_3$=1 $J_{K_{a},K_{c}}$=5$_{7,2}$--6$_{6,1}$  & compact & 70$\times$70 & 220 & 60 & \\                      
SiO $v$=9 $J$=17--16  & compact & 60$\times$60 & 360 & 170&  \\                            
U692727     & -- &    & 120 & 45 & low S/N \\                                                               
$^{13}$CS $J$=15--14     & compact & 50$\times$50 & -200 & -70 & absorption \\             
SiO $v$=0 $J$=16--15  & extended+ring & 300$\times$110 & 600 & 220 & absorption \\    

  \end{tabular}
  \label{tab.data.b9}
\end{table*}

\section{Analysis}\label{sec.ana}

\subsection{Continuum maps}\label{sec.ana.cont}

Our low angular resolution continuum map at 1.3\mm{} allows for the study of the most extended continuum emission in \raqr{} (see \fig\ref{fig.map.conB6Low}). In this map, we can identify an intense and compact central component. This component is dominated by the continuum emission of the radio-photosphere of the Mira component and its close surroundings (see below). The flux of this component integrated over the beam is $\sim$\,40\mJy. Extended emissions appear in the northeast-southwest direction. We identified these emissions with the jet, which is also seen in lower frequencies and optical images. Secondary maxima located at offsets ($-$0\secp123,$-$0\secp33), and ($-$0\secp73,$-$0\secp70), with intensities of $\sim$4 and $\sim$0.57\mJyBeam{}, respectively, are part of the innermost jet clouds previously detected in H$\alpha$ \citep{schmid17}.

The high angular resolution continuum maps at 1.3 and 0.45\mm{} also show an intense compact component in the central position (see \fig\ref{fig.map.conB6B9High}). The flux of this central component is 30 and 220\mJy{} at 1.3 and 0.45\mm{}, respectively. These values are somewhat larger than the expected photospheric flux \citep[16 and 140\mJy{},][]{bujarrabal18} and probably include dust emission of nearby shells. 
 
By comparing the intensities measured for the compact central component of the high-resolution continuum maps at 1.3 and 0.45\mm, we estimated a spectral index of 1.88. Therefore, the nature of this region is compatible with the extended radio photosphere of an AGB star \citep[see][]{vlemmings19,vlemmings15,planesas16}.  


Thanks to the very high spatial resolution of our observations, we could study the innermost regions of the \raqr{} system, where the jet is formed. This jet appears as an extended emission along the northeast-southwest direction, as was the case in the low-resolution image (see \figs\ref{fig.map.conB6Low} and \ref{fig.map.conB6B9High}). We therefore conclude that the orientation of the jet does not significantly change from arcsecond to milliarcsecond scales. In the particular case of the high-resolution map at 1.3\mm{} (Band\,6), the continuum emission extends toward the predicted position of the WD (blue symbol in \fig\ref{fig.map.conB6B9High}). Following the arguments in \cite{bujarrabal18}, this extension would be due to the emission of the ionized gas surrounding the WD, including the accretion disk and the base of the jet. The coincidence with the predicted WD position confirms the predicted orbital parameters and binary movements (Sect.\ref{sec.introduction}).

We found that the continuum extended emissions are very similar at 0.9 and 1.3\mm{} in the central region. For distances larger than 1\arcsec{}, more extended structures appear at 1.3\mm, mostly due to the better sensitivity achieved.

\subsection{H30$\alpha$ maps}\label{sec.ana.recom}

Maps per velocity channel of H30$\alpha$ can be seen in \figs\ref{fig.map.b6_h30a} and \ref{fig.map.b6_h30a_lowres} for the high- and low-resolution data, respectively. Due to the better sensitivity of the low-resolution map, the structures detected are wider. The high-resolution maps provide precise information about the brightness distribution at the innermost regions of the system. To show the brightness distribution of this line better, we used a map size of 0\secp30\,\x0\secp15 and 3\secp0\,\x1\secp5 in \figs\ref{fig.map.b6_h30a} and \ref{fig.map.b6_h30a_lowres}, respectively. 

The emission of H30$\alpha$ extends in the direction northeast-southwest for both high- and low-resolution maps. This emission is compatible with the jet emitted from the accretion disk formed around the WD. This atomic line spans a velocity range between --80.6 and 29.8\kms{}, with its maximum intensity between --37.4 and --27.8\kms{} in the regions close to the center. Therefore, we detected the projected velocity of the jet spanning $\pm$55\kms{} with respect to the systemic velocity, which is much larger than the range found for the molecular lines but rather low for a jet emerging from a WD. In \raqr{} and other symbiotic systems showing jets, their velocity is $\sim$1000\kms{} and comparable to the escape velocity from the WD surface \citep[see e.g.][]{kellogg07,schmid17,galan22}.  This may indicate that the jet is significantly inclined with respect to the orbital plane (i.e., almost parallel to the plane of the sky). \cite{kellogg07} reported a tangential velocity of 768$\pm$40\kms{} (corrected for our adopted distance) in the NE outer lobe of the jet observed in X-rays at distances larger than 5\arcsec{}, a region that is not detected in our data. 

\fig\ref{fig.map.conB6H30aLowHigh} shows the emissions of H30$\alpha$ and continuum at 1.3\mm{} together for the high- and low-resolution configurations. In the low-resolution maps, the brightness distribution of H30$\alpha$ and the continuum are coincident, and both emissions describe the structure of the jet at the arcsecond scale. This image of H30$\alpha$ is similar to the distribution of H$\alpha$ in the optical \citep{schmid17}. The remarkable coincidence between the extended continuum, H30$\alpha$, and H$\alpha$ distributions indicates that this wide continuum component is due to emission from the jet ionized gas. As mentioned, the continuum of the primary is an important component of the total flux in the high-resolution maps at 1.3\mm{}. This can explain why the emission of H30$\alpha$ is slightly displaced to the west of the continuum emission (see bottom panel of \fig\ref{fig.map.conB6H30aLowHigh}), as the RRL emission is associated with the WD surroundings and the jet, not with the radio phothosphere of the AGB star.

\subsection{Molecular line maps}\label{sec.ana.mol}

The \doce{} transitions (see Figs.\,\ref{fig.map.b6_12co21}, \ref{fig.map.b7_12co_j3_2}, and \ref{fig.map.b9_co_j6_5}) show the most extended structures, which are almost 1\secp0\,\x0\secp3 in the R.A. and Dec. directions, respectively ($\sim$260\,$\times$\,80 AU, 3.9·10$^{15}$\,$\times$\,1.2·10$^{15}$\cm{} for the adopted distance). The structure is elongated almost in the direction of the projected orbit of the system. For the rotational transitions $J$= 2--1, 3--2, and 6--5 of \doce{}, the emission presents a maximum intensity of 250, 400, and 590\,K, respectively. These lines show the most intense thermal emission detected in \raqr. %

The emission of the isotopic substitution \trece{} is more compact, showing an extension of 300\,\x200 and 60\,\x30\mas{} for the transitions $J$=2--1 and $J$=3--2 (see Figs.\,\ref{fig.map.b6_13co21} and \ref{fig.map.b7_13co_j3_2}). The differences noticed between the size of \trece{} at 1.3 and 0.9\mm{} can be explained by the higher sensitivity achieved at 1.3\mm{}. 

Many transitions of \sio{} tend to form rings around the central object (see Figs.\,\ref{fig.map.b7_29sio_v0_j8_7}, \ref{fig.map.b9_sio_v0_j16_15}, and \ref{fig.map.b9_sio_v1_j16_15}). This structure is usually detected in Mira-type variables \citep[see][]{kaminski16}. In our maps, the rings are 90\mas{} wide and centered around the adopted central position, which is supposed to be coincident with the AGB star (see Sect.\,\ref{sec.observations}). The emission in these cases also tends to extend mostly in the east-west direction. Ring-like or central minimum structures are also found in the \co{} $J$=3--2 and $J$=6--5 transitions within the characteristic wide structure. We remind that we are presenting continuum-subtracted line maps, \sect\ref{sec.observations}; however, we think that such a process is not enough to explain the observed central minima (see discussion in \sects\ref{sec.ana.obser} and \ref{sec.models}), and therefore, those structures are not artifacts.

We observed a double-peaked or "peanut-like" structure in some molecular transitions, for example in the \so{} transitions (see Figs.\,\ref{fig.map.b9_so_v0_17_16} and \ref{fig.map.b9_so_v0_j1616_1515}). In these cases, the emission is more compact than for \co{} and \sio{}. The two peaks are located at 10\,mas in the northwest and southeast directions and are almost perpendicular to that of the continuum and RRLs. We find that the ring- and peanut-like shapes are often difficult to distinguish, as seen in the brightness distribution of the \sio{} $v$=1 $J$=16-15 and the \so{} $^3\Sigma$ $v$=0 $N_J$=17$_{16}$--16$_{15}$ lines (see Fig.\,\ref{fig.map.b9_sio_v1_j16_15} and \ref{fig.map.b9_so_v0_17_16}). We think that the peanut-like shape could be an incomplete ring-like distribution.

Single-peak compact distributions are clearly shown by some molecular transitions (see, e.g., Figs.\,\ref{fig.map.b7_h2o_v2_2} and \ref{fig.map.b7_so_v1_j9-8}). The deconvolved size of the compact structure is between 10 to 15 mas, depending on the molecule. We also found that these peaks are usually not centered on top of our adopted central position (see, e.g., Fig.\,\ref{fig.map.b7_h2o_v2_2}). In these cases, the spatial distribution could be a peanut-like one in which the southeast peak is not detected.

Maser emission is clearly detected in the $^{28}$\sio{} $v$=0,1 $J$=5--4 lines (Fig.\,\ref{fig.map.b6_sio_v0_j5_4} and \ref{fig.map.b6_sio_v1_j5_4}), and the $v$=1 line is particularly strong. This line shows at least three maser spots located $\sim$15\mas{} around the center of the map. For an assumed distance to the object of 265\,pc, we found that the distance from the center to the maser spots is $\sim$\,6·10$^{13}$\cm{} ($\sim$4\,AU, a few stellar radii from the central star). Similar structures are usually detected in the same lines from AGB stars, including VLBI images of \raqr{} \citep[see][]{diamond94,soriaruiz04,cotton04,cotton06,ragland08,kamohara10}. These lines show extended structures in the central velocities between --26.2 and --19.8\kms{} and --32.6 and --26.2\kms{} for the $v$=0 and $v$=1 transitions, respectively. The weak extended emissions in these maps are probably of thermal origin rather than maser, as they are similar to those seen in other \sio{} transitions (see \figs\ref{fig.map.b7_29sio_v0_j8_7}, \ref{fig.map.b9_sio_v0_j16_15}, and \ref{fig.map.b9_sio_v1_j16_15}).

Some molecular species show absorption at positive velocities between $-$16.6 and $-$7.0\kms{} (see, e.g., several \sio{} transitions; Figs.\,\ref{fig.map.b7_29sio_v0_j8_7}, \ref{fig.map.b9_sio_v0_j16_15}, and \ref{fig.map.b9_sio_v1_j16_15}). We also found an almost pure absorption in the $^{13}$CS $J$=15--14 line (Fig.\,\ref{fig.map.b9_C13S}). Similar absorption features have been found in AGB stars, such as R\,Leo and Mira, whose spectra also show red-shifted absorption in front of the star position \citep{fonfria19,kaminski17}. These spectral features were proposed to be the result of radial movements of the photosphere layers during the stellar pulsation of the AGB star \citep{hinkle79,freytag23}. We think that the absorption found in these lines traces relatively cool gas falling back to the AGB component and partially absorbing the stellar continuum. The presence of in-falling gas after the interaction with the WD was predicted by hydrodynamical models of this system \citep[see][]{bujarrabal21,freytag23}. 

\begin{figure}[htpb]
  \centering{\resizebox{9.0cm}{!}{
  \includegraphics{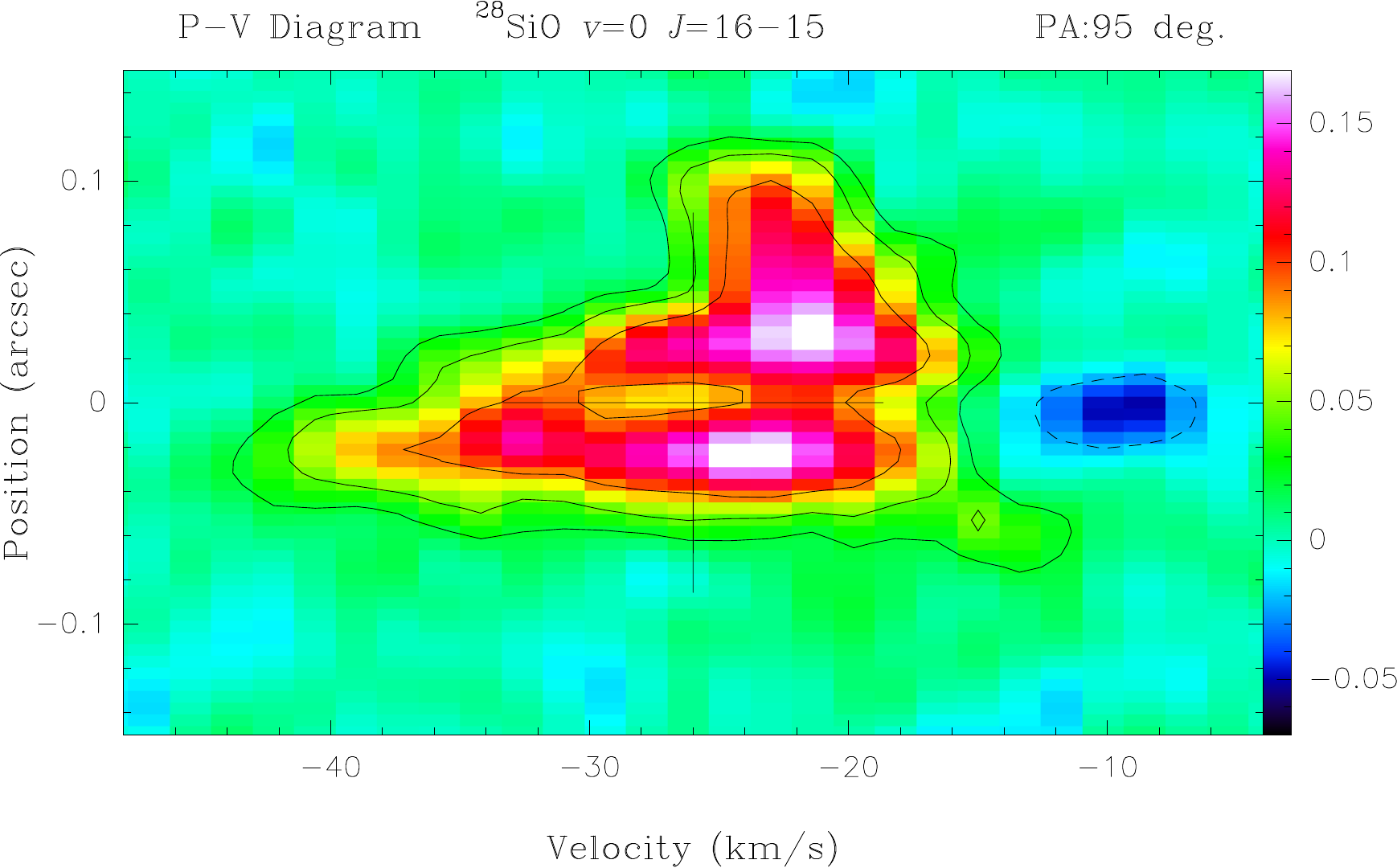}}}
  \caption{Position-velocity map along the orbital direction for the $^{28}$SiO $v$=0 $J$=16--15 line. Negative position offsets are westward. The cross marks the position of the continuum peak (origin of coordinates) and the adopted systemic velocity. The brightness units are \JyBeam. The contours are logarithmic, with a step of a factor of two and a first level of $\pm$20\mJyBeam. The dashed contours represent negative values.}
    \label{fig.pv.sio_v0_16_15}
\end{figure}

Figure\,\ref{fig.pv.sio_v0_16_15} shows the position-velocity diagram of $^{28}$SiO $v$=0 $J$=16--15 for a PA of 95\,º, corresponding to the projected shape of the orbit (see \sect\ref{sec.introduction}). We can identify two relative minima in the central position at the central velocity, $-$27\kms{}, and at $-$9\kms. As mentioned before, the in-falling gas to the AGB causes the second absorption around --9\kms. We detected gas with negative velocities with respect to the systemic velocity of \raqr{}, around $-$37\kms, at $-$0\secp02. This can be understood as material approaching the observer. In addition, at +0\secp02 there is gas receding from us at relative positive velocities around $-$22\kms. In both cases, these movements are consistent with the existence of gas close to the primary being dragged by orbital movement of the companion. The feature at $-$0\secp03 and $-$23\kms may correspond to a brightness enhancement in the general ring-like structure.

\subsection{Observational effects on the final brightness distribution}\label{sec.ana.obser}

In this section, we analyze the effects introduced by the continuum subtraction in the observed spatial brightness distribution of the molecular lines and the impact of the non-circularity of the restoring beams in our line maps. Firstly, molecular emission can come from clumpy structures, but the current information prevents confirming this. Therefore, we assumed that the emission of the molecules studied here is smooth and homogeneous. As a result, the brightness distributions are considered the average emission of every possible emitting clump. 

To examine the brightness distribution shown by the different lines, we subtracted the continuum emission. As a result, our maps show the emission line excess with respect to the continuum level. This procedure can affect the observed brightness distribution where the continuum is intense, and the line is optically thick. However, this is necessary to obtain information about the size and shape shown by the less intense lines.   
We used the emission of three central velocity channels of five molecules to compare the spatial brightness distribution before and after continuum subtraction. \fig\ref{fig.map.centralVelocitiesCont} shows the emission of the central velocity channels of several molecules before the continuum emission is removed in the \textit{uv} tables. The spatial brightness distributions shown by the abundant molecules, such as \doce{} $J$=6--5 and $^{28}$\sio{} $v$=0 $J$=16--15, are extended with a compact maximum in the central position, which is dominated by the continuum emission (see Sec.\,\ref{sec.ana.cont}). However, the central channels of the less abundant molecules (e.g., SO $^3\Sigma$ $v$=0 $N_J$=16$_{16}$--15$_{15}$, H$_2$O $v_3$=1 $J_{K_{a},K_{c}}$=5$_{7,2}$--6$_{6,1}$, and SiO $v$=9 $J$=17--16,) show only a central component. In these cases, distinguishing the pure line emission is difficult, and we could not estimate the size of their distributions. 

\figs\ref{fig.map.centralVelocitiesCont} and \ref{fig.map.centralVelocities} show the same velocity channels of the lines but in \fig\ref{fig.map.centralVelocities} the continuum distribution is subtracted. The brightness distributions exhibited by \doce{} $J$=6--5 and $^{28}$\sio{} $v$=0 $J$=16--15 in outer regions are similar to those shown in \fig\ref{fig.map.centralVelocitiesCont} without removing the continuum. However, because the \doce{} and $^{28}$\sio{} molecules are optically thick, the central positions are strongly affected by the continuum subtraction. As a result, after the continuum is removed a relative minimum appears at the center.

For the most compact and weak lines, we could only see the line emission shape after removing the contribution of the continuum. In these cases, no absorption features were detected, suggesting that these lines are optically thin or optically thick but with a small filling factor. In the intermediate cases (e.g., SO $^3\Sigma$ $v$=0 $N_J$=17$_{16}$--16$_{15}$), we could identify some extended emission after continuum subtraction, with a moderate central minima.

We note that the final brightness distributions can also be affected by the beam shape. The convolution of a symmetric ring with an elongated beam gives a non-symmetric distribution, with two maxima parallel to the beam elongation, and eventually generates spurious peanut-like shapes. In our cases, such an effect is expected to be minor (\sect\ref{sec.models}).

\section{Modeling the compact component of the molecular emission}\label{sec.models}

We have developed a simple model to give a general description of the brightness distributions of molecular emission in \raqr{} (see \sects\ref{sec.results.lines} and \ref{sec.ana.mol}). The distribution of \co{}, the most extended emission in this object, was successfully reproduced by \cite{bujarrabal21} using complex hydrodynamical and photodissociation models. \cite{bujarrabal21} argue that the final picture of the \co{} in \raqr{} is the result of the interaction between the AGB and the gravitational pull of the WD. The intense UV emission from the photosphere of the hot companion plays an important role in the distribution of molecular content in this system. \cite{bujarrabal21} found that this UV emission efficiently photodissociates molecules, and therefore, only the molecules in the densest regions survive thanks to the stronger shielding there. In this work, we focus on the compact emission detected in molecules such \sio{}, \so{}, and others, which are much less abundant than \co{} and can therefore only survive in regions closest to the AGB star. Our code is similar to those extensively used and described previously to model emission of gas in rotation and expansion in evolved objects \cite[see e.g.][]{gallardo21,bujarrabal17,bujarrabal21}. We show the synthetic maps derived from our best-fit model for the particular case of \sio{} emission. However, this model aims to be a general description to understand the distribution of less abundant molecules than \co{}. 

We assumed local thermodynamic equilibrium (LTE) to obtain the coefficients of absorption and emission in each point of the circumstellar shells. As we explain below, our best-fit model gives density values of about 10$^9$\,cm$^{-3}$, which is above the critical density of the considered species. Therefore, the excitation temperatures are well described by a rotational temperature that in our cases is similar to the kinetic temperature.

To simulate the molecular emission, we used ray tracing with a high number of lines of view that are convolved with the observational beam. Our model nebula consists of an oblate ellipsoidal shell in radial expansion for the emission and an in-falling spherical shell for the absorption (see \fig\ref{fig.map.b9_sio_v0_j16_15}). On the one hand, the size of the ellipsoidal shell is defined by $R_{o.equ}$ and $R_{o.axi}$, which are the outer radii in the equatorial and axial directions of the nebula, respectively. The inner radii of this component are parametrized by $R_{i.equ}$ and $R_{i.axi}$. On the other hand, the origin of the absorption detected in the central region of our maps at positive velocities (see \sect\ref{sec.ana.mol}) is unclear, and hence, to model this component, we have to take some assumption about its geometry and physical conditions. To simplify our calculations, and since the existing  information on this component is poor, we assumed a spherical shell of outer and inner radii R$_{o.equ}$ and R$_{i.equ}$ (therefore, $R_{o.equ}$ = $R_{o.axi}$ and $R_{i.equ}$ = $R_{i.axi}$ for the in-falling shell). We stress that in-falling gas in regions very close to the AGB star is predicted by hydrodynamical models in our case. 

Since the material in these shells is located in the innermost layers of the envelope around the AGB star, where the dominant movements are very complex, we assumed a constant radial velocity to simplify our calculations. We checked several laws for the velocity as a function of the distance, but the final results are very similar to the constant velocity assumption. Since we did not detect rotating gas in the central region of our maps, where our model properly works, we have not included tangential velocities.

\begin{figure*}[htpb]
  \centering{\resizebox{18.2cm}{!}{
  \includegraphics{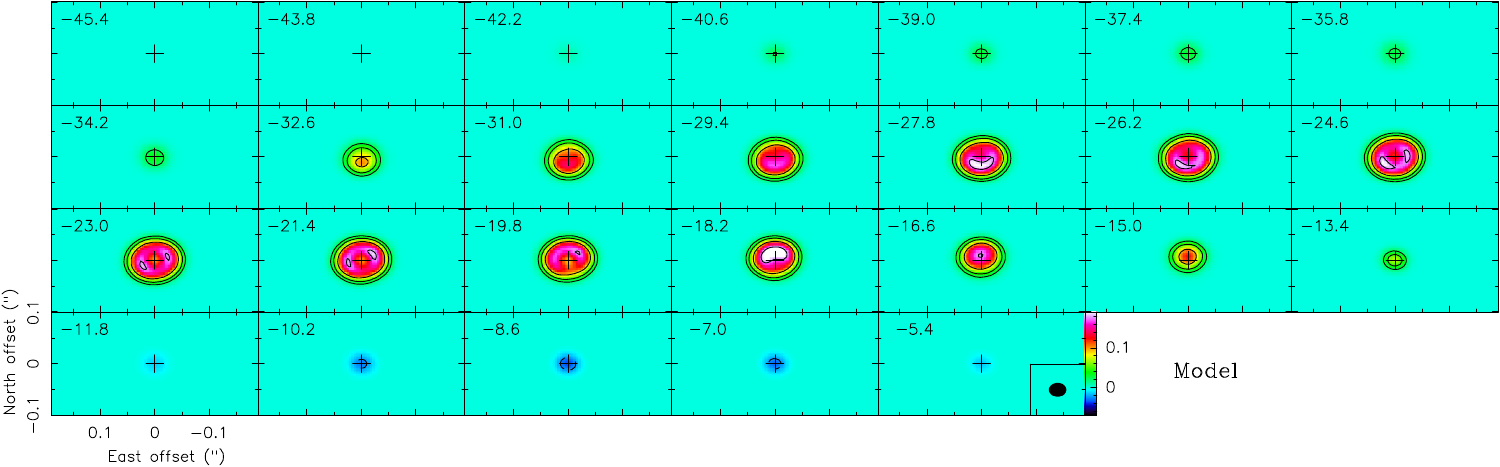}}}
  \caption{Synthetic maps per velocity channel obtained using our model for the $^{28}$SiO $v$=0 $J$=16--15 line. The LSR velocities are indicated in the upper-left corners. The brightness units are \JyBeam. The contours are logarithmic, with a step of a factor of two and a first level of $\pm$20\mJyBeam. The dashed contours represent negative values. The HPBW is shown in the inset in the last panel.}
    \label{fig.map.model}
\end{figure*}

We used the following laws for the density ($\rho$) and temperature ($T$):

\begin{gather}
         \rho\left(r_{equ},r_{axi}\right) =  \rho_{o}\left(\frac{1}{r_n}\right)^{\,k_{\rho}}\label{eq.law.den} ,\\
         T\left(r_{equ},r_{axi}\right) =  T_{o}\left(\frac{1}{r_n}\right)^{\,k_{T}} \label{eq.law.tem} ,
\end{gather}

\noindent where $\rho_0$ and $T_0$ are the temperature and density in the outer limits of the shells; $k_{\rho}$ and $k_{T}$ are the slopes of the potential laws for the density and temperature, respectively; and $r_n$ is a normalized dimensionless variable defined as follows:

\begin{gather}
 r_n =  \sqrt{\left(\frac{r_{equ}}{R_{o.equ}}\right)^{2}+\left(\frac{r_{axi}}{R_{o.axi}}\right)^{2}} \label{eq.rad}.
\end{gather}

\noindent Here, $r_{equ}$ and $r_{axi}$ are the distances to the equator and axis of the shell, respectively, for the considered point. 

As explained before (see \sect\ref{sec.ana.obser}), the effects introduced by the continuum subtraction in the final brightness distributions seen in the maps of the molecular lines must be taken into account. To treat the continuum sources in the simulation of the resulting line emission, we assumed two continuum components. A central compact photosphere that is completely opaque and has the extent and temperature described in \cite{bujarrabal18}; the molecules placed between the photosphere and the observer absorb its emission and add their line emission. We also included an extended dust component. Since the structure of the dust distribution and temperature are not well known, we took a simple Gaussian distribution for the dust emission and absorption coefficients. We assumed the dust to be optically thin, which is justified at the relatively low frequencies discussed here. In regions where dust and molecules coexist, the  transfer treatment must consider that the total emission and absorption coefficients ($j_t$, $k_t$) are locally the addition of the coefficients of the dust and molecules:

\begin{gather}
j_t = j_d + j_m \label{eq.jt},\\
k_t = k_d + k_m \label{eq.kt},
\end{gather}
where the subscripts \textit{d} and \textit{m} stand for dust and molecules, respectively. In these circumstances, the total source function satisfies:
\begin{gather}
S_t = \frac{S_d k_d + S_m k_m}{k_d + k_m} \label{eq.st}.
\end{gather}
The assumption of optically thin dust simplifies much of the implementation of dust in the numerical treatment of radiative transfer. The dust distribution was normalized to match the observations of continuum emission alone (out of the frequency ranges of the molecular lines). Finally, the predicted continuum brightness distribution was subtracted from the resulting dust plus photosphere plus line synthetic maps, which we calculated under the above-described scenario. In this way, we simulated the actual line observations in which the continuum distribution is subtracted. 

As mentioned (\sect\ref{sec.ana.obser}), for strongly elongated beams and shell-like emitters, convolution may generate a couple of spurious maxima. However, our modeling shows that our maps, derived using nearly circular beams (see \tabs\ref{tab.lines.b6}, \ref{tab.lines.b7}, and \ref{tab.lines.b9}), practically do not show such an effect, and the convolved maps properly reflect the actual brightness distribution. Nevertheless, the continuum subtraction may cause a central minimum in the final maps, \sect\ref{sec.ana.obser}, so we numerically checked that our model nebula must include a central minimum of density to replicate the observations.

\begin{table}[htpb]
\caption{Best-fit model parameters.}
\begin{tabular}{ l c l }
\hline \noalign{\smallskip} 
  Parameter   & Expanding shell  & In-falling component   \\
\noalign{\smallskip}
\hline \noalign{\smallskip}\smallskip

 R$_{o.equ}$ (cm)  & 1.6$\times$10$^{14}$ &   6$\times$10$^{13}$ \\  \smallskip 
 R$_{o.axi}$ (cm)  & 1.1$\times$10$^{14}$ &   6$\times$10$^{13}$  \\ \smallskip
 R$_{i.equ}$ (cm)  & 9.6$\times$10$^{13}$ &   3.6$\times$10$^{13}$ \\  \smallskip 
 R$_{i.axi}$ (cm)  & 6.6$\times$10$^{13}$ &   3.6$\times$10$^{13}$ \\ \smallskip
 $\rho_0$ (cm$^{-3}$) & 5$\times$10$^{8}$ &   4$\times$10$^{9}$  \\ \smallskip
 $T_0$ (K) & 700 & 300 \\  \smallskip
 $k_{\rho}$ & 0.5 & 0 \\ \smallskip
 $k_{T}$    & 1.0 & 0 \\ \smallskip     
 $V_{exp}$ (\kms) & 9.0 & -17.0 \\ \smallskip                                           
  \end{tabular}
  \label{tab.model.para}
\end{table}

As mentioned before, we used the emission of the \sio{} $v$=0 $J$=16--15 line (see \fig\ref{fig.map.b9_sio_v0_j16_15}) as an example to show the results obtained using our model. For $^{28}$SiO, we assumed a relative abundance of 6.6$\cdot10^{-6}$, which is compatible with the values found in O-rich Mira-type variables \citep[see][]{massalkhi20}. We found the best-fit model by manually varying the parameters to adjust the synthetic maps to the observations. \tab\ref{tab.model.para} shows the parameters derived from the best-fit model. We found radii $\sim$10$^{14}$\cm{} (25\,mas projected in the sky for a distance of 265\,pc). We note that these sizes are smaller than the semi-major axis of the orbit (see \sect\ref{sec.introduction}), and therefore, we are observing the molecular emission in intraorbital regions. The outer radii of the expanding shell are well determined since our maps provide a direct image of the external structure of the nebula. By varying the width of this shell, we found a value of $\sim$7$\cdot10^{13}$\cm{} for our best-fit model. Our best-fit model shows that the in-falling component is constrained by the width of the expanding shell and by the photosphere of the AGB star, whose radius is $\sim$3$\cdot10^{13}$\cm{}. However, the uncertainty of this geometry is high, and the absorption is also compatible with a component more extended than the expanding shell. We estimated that the uncertainty in the determination of the sizes of the expanding shell and in-falling component is smaller than 20\% and $\sim$50\%, respectively. Therefore, the obtained result concerning the size of the in-falling component is not conclusive. When comparing the model predictions with our maps per velocity channels, we accurately derived velocities of 9 and $-$17\kms{} for the expanding and in-falling components, respectively.

From the best-fit model, we find that the projection of the expanding shell is inclined at a PA of 105\tdeg, which is similar to the orientation of the orbit and of the equatorial component of the large-scale optical nebula (see \sect\ref{sec.introduction}). This confirms the link between the structures of these inner shells with the gravitational pull of the companion. 

We find that the in-falling component is cooler and denser than the expanding shell. Since we do not have precise information about this component, we assumed that the temperature and density are constant, that is, the slopes of the density and temperature laws (see Eqs.\,(\ref{eq.law.den}) and (\ref{eq.law.tem})) for the in-falling component are zero.

\fig\ref{fig.map.model} shows the synthetic maps per velocity channel derived from the best-fit model, which replicates the brightness distributions found in the maps of the \sio{} $v$=0 $J$=16--15 line (see \fig\ref{fig.map.b9_sio_v0_j16_15}). In particular, our model predicts the ring-like shape seen in the maps of this line, and it is particularly consistent for the central velocities, between $-$25 and $-$18\kms. The synthetic maps also reproduce the absorption seen in the positive velocity channels of this line. Our model also reproduces, qualitatively, the brightness distributions seen in other lines, such as SO $^3\Sigma$ $v$=0 $N_J$=17$_{16}$--16$_{15}$ and $^{28}$SiO $v$=4 $J$=16--15 (see \figs\ref{fig.map.b9_so_v0_17_16} and \ref{fig.map.b9_SiOv4_16_15}), including the rings and peanut-like shapes seen in the maps of these and other lines. 

The brightness distributions seen in other molecular lines that show compact emission, such as the SO $^3\Sigma$ $v$=1 $N_J$=9$_8$--8$_7$ and H$_2$O $v_2$=2 $J_{K_{a},K_{c}}$=3$_{2,1}$--4$_{1,4}$ lines (see \figs\ref{fig.map.b7_so_v1_j9-8} and \ref{fig.map.b7_h2o_v2_2}), can also be explained using our model. In these cases, we must assume that the outer radii are smaller than those for the \sio{} $v=0$ $J$=16--15 line. These are even less abundant species, and as a result, they can only survive in regions even closer to the AGB star, where the shielding is stronger.

Significantly more extended emission cannot be predicted by our model because they come from more complex regions where the dynamics cannot be described using a simple expanding shell. This is the case of the emission located in the southeast and southwest direction at $-$29.4 and $-$23\kms, respectively, in the maps of the $^{28}$SiO $v$=0 $J$=16--15 line (see \fig\ref{fig.map.b9_sio_v0_j16_15}). We relate these features with the innermost part of the components labeled B and "?" of the $^{12}$CO $J$=3--2 emission described by \cite{bujarrabal21}. Other extended emission identified by \cite{bujarrabal21} in CO, such as components C, D, and W, have no counterpart in the \sio{} maps; \sio{} seems to be tracing the innermost regions of the equatorial structures seen in \co{} and identified as spiral arcs. Other transitions of $^{28}$SiO, $^{29}$SiO, $^{13}$CO, and SO also show extended emissions that are coincident in velocity and position with the emission and \co{}. For example, one can see extended emission toward the southeast direction at $-$23\kms{} in the maps of the $^{13}$CO $J$=3--2 and SO $^3\Sigma$ $v$=0 $N_J$=17$_{16}$--16$_{15}$ lines (see \figs\ref{fig.map.b7_13co_j3_2} and \ref{fig.map.b9_so_v0_17_16}). 

Our modeling of the compact emission in various molecular lines agrees with the general trends of chemical model predictions \citep{bujarrabal21}. The photo-induced chemistry in \raqr{} favors the survival of molecules in the orbital plane. In addition, most of the molecular content must be close to the photosphere of the AGB star, especially for the less abundant molecules whose self-shielding is negligible. In summary, the geometry of the model shells described before is qualitatively consistent with those predicted by the models of chemistry in stellar systems \citep{bujarrabal21}.

\section{Conclusions }\label{sec.con}

The ALMA observations of \raqr{} in Bands~6,~7, and\,~9 (1.3, 0.9, and 0.45\mm{}) have been analyzed in detail. The continuum emission and many lines have been mapped using different array configurations as well as natural and robust weightings in order to explore different spatial scales. The compact molecular emission seen in the central region of our maps has been modeled using simple shell components and considering the effects of the continuum subtraction. Our conclusions are summarized as follows:
\begin{itemize}

\item Radio continuum emission at 1.3 and 0.45\mm{} has been studied. The low-resolution map of the continuum at 1.3\mm{} (see \fig\ref{fig.map.conB6Low}) shows the emission of the stellar components and their surroundings at the central position as well as the bipolar jet tracing structures similar to those previously found in other wavelengths \citep[see e.g.][]{hollis97,schmid17}. 

\item High-resolution continuum maps show emission of the radio photosphere of the AGB and its closest surroundings in the central region of the image (see \fig\ref{fig.map.conB6B9High}). The extended emission shows the innermost part of the jet. At 1.3\mm{}, the brightness distribution seems to be caused by the emission of both stellar components, but due to the fact that the separation in the sky is smaller than the spatial resolution, this cannot be clearly confirmed.

\item The brightness distributions of the radio recombination line H30$\alpha$ (see \figs\ref{fig.map.b6_h30a} and \ref{fig.map.b6_h30a_lowres}) show similar structures at large scale to those seen in our continuum maps (see \fig\ref{fig.map.conB6H30aLowHigh}). Not surprisingly, the high-resolution map of H30$\alpha$ (bottom panel in \fig\ref{fig.map.conB6H30aLowHigh}) shows that the emission of atomic gas is not centered on the continuum emission since H30$\alpha$ emission is not observed near the photospheres of AGB stars. The distribution of atomic gas suggests that, as expected, the jet emerges from the region close to the companion.

\item The emission \co{} and \sio{} (thermal lines), the most abundant molecules detected in this work, show extended structures elongated in the direction of the orbital plane. The extended structures in the \co{} emission were previously identified with spiral arcs \citep{bujarrabal21}. Some components in the emission of the $^{12}$CO can also be found in the \sio{} maps (see \figs\ref{fig.map.centralVelocitiesCont} and \ref{fig.map.centralVelocities}), and therefore, the extended emission of \sio{} probably comes from the inner spiral structures. 

\item Less abundant molecules, such as \so{} and H$_2$O, show compact brightness distributions around the AGB star. We confirm that these molecules only survive in the region close to the AGB star, where the shielding is expected to be the strongest due to the high density of dust and molecules. 

\item We have presented a simple model of the inner molecular shells that provides a general explanation for the brightness distributions seen in most of the detected molecular lines. In particular, we have reproduced the emission of SiO from the intra-orbital regions. Our model consists of an oblate expanding shell and a yet inner spherical in-falling component. Our best-fit model provides external sizes of the shells $\sim$10$^{14}$\cm{} that are comparable to the orbital size. Therefore, we have mainly studied the molecular content within the orbit of the companion. 

\item We have also successfully reproduced the absorption detected in the SiO maps at positive velocities in the central position (see \figs \ref{fig.map.b7_29sio_v0_j8_7} and \ref{fig.map.b9_sio_v0_j16_15}) using a cool in-falling innermost component. Due to the uncertainty in the geometry of this component derived from the best-fit model, the in-falling material could also be located in outer regions with a size larger than the expanding shell. Taking into account the uncertainties, our best-fit model is consistent with hydrodynamical models that predict that previously ejected material may fall back to the AGB star \citep{bujarrabal21,freytag23}. Since the information about the component that causes the absorption is poor, we cannot rule out that the radial movements of the photosphere layers yield this spectral feature.

\item The chemical description by \cite{bujarrabal21} qualitatively agrees with the observational results explained in this work. Due to the intense UV radiation emitted by the WD, very rare molecules with relative abundances smaller than $10^{-6}$ only survive in regions very close to the AGB star. The self-shielding in these molecules is small, and as a result, they are easily photodissociated. Maps of more abundant molecules, including SiO and SO, show some extended structures. In these intermediate cases, with relative abundances between $10^{-6}$ and $10^{-4}$, self-shielding is higher, and the molecules can survive at distances farther from the mass-losing Mira. The most abundant molecule detected in our observations, CO, exhibits even larger structures. In all cases, the molecule-rich components tend to be confined to equatorial regions. 

\item The structures detected in the central regions of the high-resolution maps of the CO and SiO lines toward the symbiotic system \raqr{} \citep[this work and][]{bujarrabal21} suggest a link between these equatorial structures and the large-scale nebula. In our opinion, the detected spiral structures lead to the formation of the incipient hourglass shape shown by material in the optical and IR pictures (see \sect\ref{sec.introduction}).

\end{itemize}

We have studied the distribution of molecular gas in the innermost region of the symbiotic system \raqr{} via high-resolution ALMA maps during its periastron. Further observations can help study the evolution of the distribution of material in time. More high-resolution maps, in different epochs, would allow for analysis of the variability of the distribution of molecular gas and the gravitational influence and the UV field of the WD in the shaping of the nebula as it travels along its orbit.

\begin{acknowledgements}
This work is part of the I+D+i project EVENTs/Nebulae Web, PID2019-105203GB-C21, funded by the Spanish MCIN/AEI/10.13039/501100011033. J.M. acknowledges the National Science Centre, Poland, grant OPUS 2017/27/B/ST9/01940. ALMA is a partnership of ESO (representing its member states), NSF (USA) and NINS (Japan), together with NRC (Canada) and NSC and ASIAA (Taiwan), in cooperation with the Republic of Chile. The Joint ALMA Observatory is operated by ESO, AUI/NRAO and NAOJ. 
\end{acknowledgements}

\newpage
\appendix

\clearpage
\newpage

\section{Maps}\label{app.maps}
Here, we present the maps of molecules with low S/N and those which show very compact brightness distributions.


\begin{figure*}[htpb!]
  \centering{\resizebox{18.2cm}{!}{
  \includegraphics{figures/mapLines/b6_13co21.eps}}}
  \caption{ALMA maps per velocity channel of $^{13}$CO $J$=2--1 emission in R Aqr. The LSR velocities are indicated in the upper-left corners. The brightness units are \JyBeam. The contours are logarithmic, with a step of a factor of two and a first level of $\pm$3\mJyBeam. The dashed contours represent negative values. The HPBW is shown in the inset in the last panel.}
    \label{fig.map.b6_13co21}
\end{figure*}


 \begin{figure*}[htpb]
  \centering{\resizebox{18.2cm}{!}{
  \includegraphics{figures/mapLines/b7_13co_j3_2.eps}}}
  \caption{Same as \fig\ref{fig.map.b6_13co21} but for $^{13}$CO $J$=3--2 using a first level of $\pm$4\mJyBeam.}
    \label{fig.map.b7_13co_j3_2}
\end{figure*}

\begin{figure*}[htpb]
  \centering{\resizebox{18.2cm}{!}{
  \includegraphics{figures/mapLines/b7_h2o_v2_2.eps}}}
  \caption{Same as \fig\ref{fig.map.b6_13co21} but for H$_2$O $v_2$=2 $J_{K_{a},K_{c}}$=3$_{2,1}$--4$_{1,4}$ using a first level of $\pm$4\mJyBeam.}
    \label{fig.map.b7_h2o_v2_2}
\end{figure*}

 \begin{figure*}[htpb]
  \centering{\resizebox{18.2cm}{!}{
  \includegraphics{figures/mapLines/b7_30sio_v3_j8_7.eps}}}
  \caption{Same as \fig\ref{fig.map.b6_13co21} but for $^{30}$SiO $v$=3 $J$=8--7 using a first level of $\pm$4\mJyBeam.}
    \label{fig.map.b7_30sio_v3_j8_7}
\end{figure*}

 \begin{figure*}[htpb]
   \centering{\resizebox{18.2cm}{!}{
   \includegraphics{figures/mapLines/b7_12co_v1_j3_2.eps}}}
   \caption{Same as \fig\ref{fig.map.b6_13co21} but for $^{12}$CO $v$=1 $J$=3--2 using a first level of $\pm$4\mJyBeam.}
     \label{fig.map.b7_12co_v1_j3_2}
 \end{figure*}

\begin{figure*}[htpb]
  \centering{\resizebox{18.2cm}{!}{
  \includegraphics{figures/mapLines/b7_so_v1_j9_8.eps}}}
  \caption{Same as \fig\ref{fig.map.b6_13co21} but for SO $^3\Sigma$ $v$=1 $N_J$=9$_8$--8$_7$ using a first level of $\pm$4\mJyBeam.}
    \label{fig.map.b7_so_v1_j9-8}
\end{figure*}

\begin{figure*}[htpb]
  \centering{\resizebox{18.2cm}{!}{
  \includegraphics{figures/mapLines/b7_so_v0_j8_7.eps}}}
  \caption{Same as \fig\ref{fig.map.b6_13co21} but for SO $v$=0 $J$=8--7 using a first level of $\pm$4\mJyBeam.}
    \label{fig.map.b7_so_v0_j8_7}
\end{figure*}

\begin{figure*}[htpb]
  \centering{\resizebox{18.2cm}{!}{
  \includegraphics{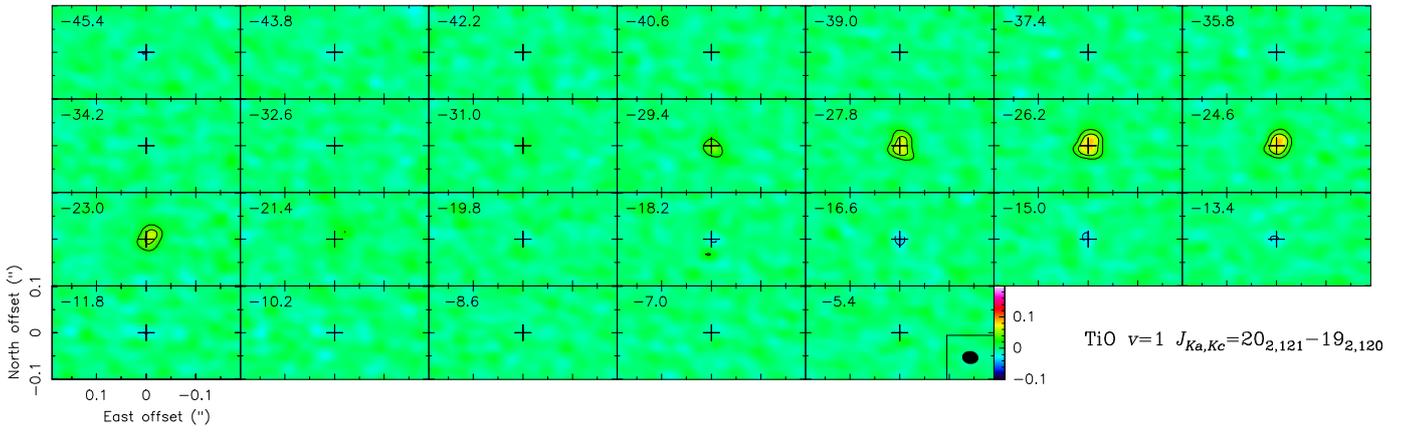}}}
  \caption{Same as \fig\ref{fig.map.b6_13co21} but for TiO $v$=1 $J_{K_{a},K_{c}}$=20$_{2,121}$--19$_{2,120}$ using a first level of $\pm$25\mJyBeam.}
    \label{fig.map.b9_TiO_v1_20_2_121_19_2_120}
\end{figure*}

\begin{figure*}[htpb]
  \centering{\resizebox{18.2cm}{!}{
  \includegraphics{figures/mapLines/b9_49TiO_20_19.eps}}}
  \caption{Same as \fig\ref{fig.map.b6_13co21} but for $^{49}$TiO $J$=20--19 using a first level of $\pm$15\mJyBeam.}
    \label{fig.map.b9_49TiO_20_19}
\end{figure*}

\begin{figure*}[htpb]
  \centering{\resizebox{18.2cm}{!}{
  \includegraphics{figures/mapLines/b9_SiOv4_16_15.eps}}}
  \caption{Same as \fig\ref{fig.map.b6_13co21} but for $^{28}$SiO $v$=4 $J$=16--15 using a first level of $\pm$20\mJyBeam.}
    \label{fig.map.b9_SiOv4_16_15}
\end{figure*}

\begin{figure*}[htpb]
  \centering{\resizebox{18.2cm}{!}{
  \includegraphics{figures/mapLines/b9_U675616.eps}}}
  \caption{Same as \fig\ref{fig.map.b6_13co21} but for U675616 using a first level of $\pm$15\mJyBeam.}
    \label{fig.map.b9_U675616}
\end{figure*}

\begin{figure*}[htpb]
  \centering{\resizebox{18.2cm}{!}{
  \includegraphics{figures/mapLines/b9_AlO_18_17.eps}}}
  \caption{Same as \fig\ref{fig.map.b6_13co21} but for AlO $N$=18--17 using a first level of $\pm$20\mJyBeam.}
    \label{fig.map.b9_AlO_18_19}
\end{figure*}

\begin{figure*}[htpb]
  \centering{\resizebox{18.2cm}{!}{
  \includegraphics{figures/mapLines/b9_U689075.eps}}}
  \caption{Same as \fig\ref{fig.map.b6_13co21} but for U689075 using a first level of $\pm$20\mJyBeam.}
    \label{fig.map.b9_U689075}
\end{figure*}

\begin{figure*}[htpb]
  \centering{\resizebox{18.2cm}{!}{
  \includegraphics{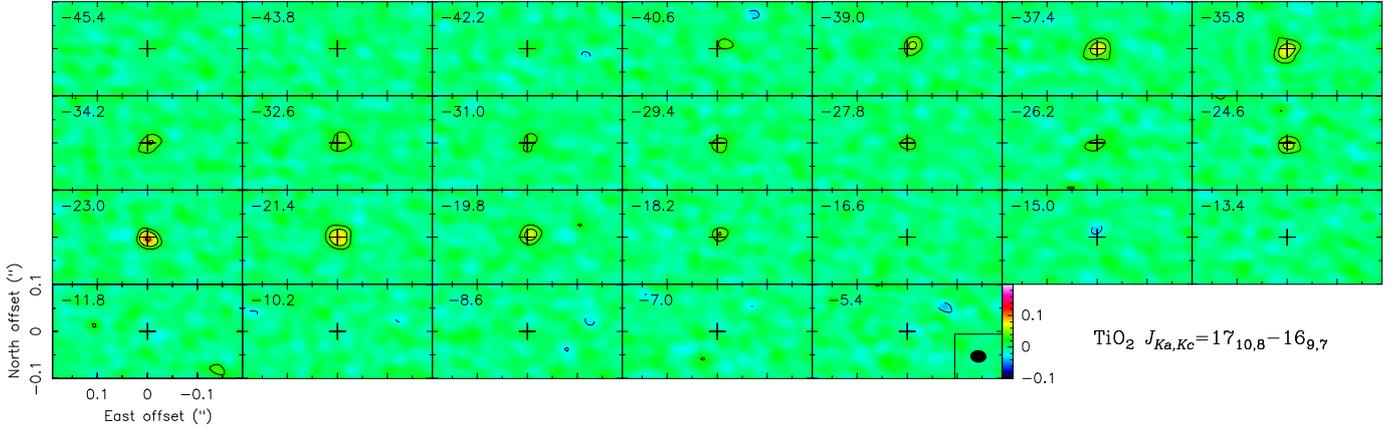}}}
  \caption{Same as \fig\ref{fig.map.b6_13co21} but for TiO$_2$ $J_{K_{a},K_{c}}$=17$_{10,8}$--16$_{9,7}$ using a first level of $\pm$25\mJyBeam. Emission of the AlF $J_{F}$=21$_{41/2}$-20$_{43/2}$ line is shown from -40.6 to -29.4\kms{} LSR.}
    \label{fig.map.b9_TiO2_17_10_8_16_9_7}
\end{figure*}

\begin{figure*}[htpb]
  \centering{\resizebox{18.2cm}{!}{
  \includegraphics{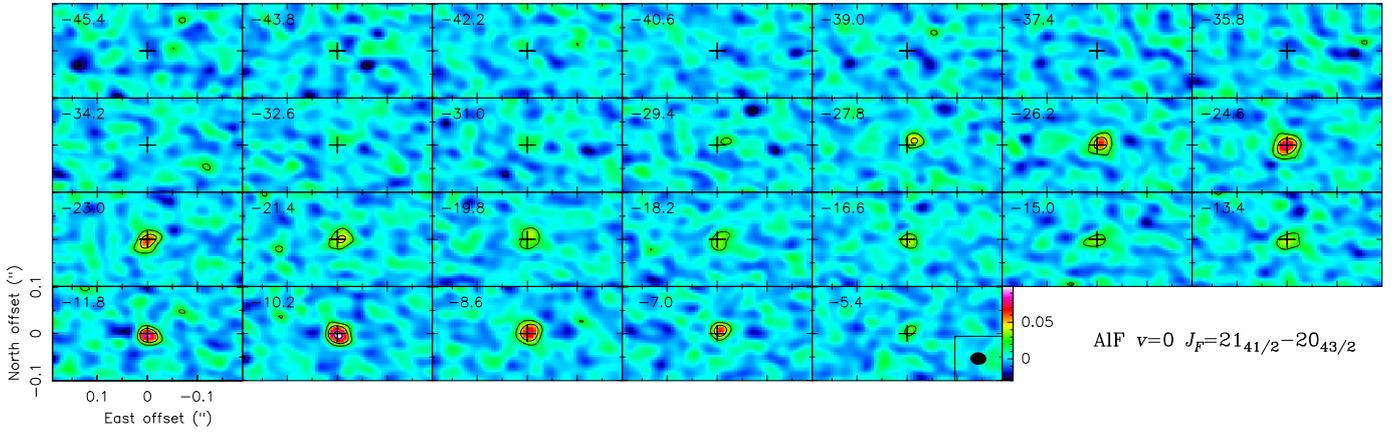}}}
  \caption{Same as \fig\ref{fig.map.b6_13co21} but for AlF $J_{F}$=21$_{41/2}$-20$_{43/2}$ using a first level of $\pm$22\mJyBeam. Emission of the TiO$_2$ $J_{K_{a},K_{c}}$=17$_{10,8}$--16$_{9,7}$ line is shown from -13.4 to -5.4\kms{} LSR.}
    \label{fig.map.b9_AlF_21_20}
\end{figure*}

\begin{figure*}[htpb]
  \centering{\resizebox{18.2cm}{!}{
  \includegraphics{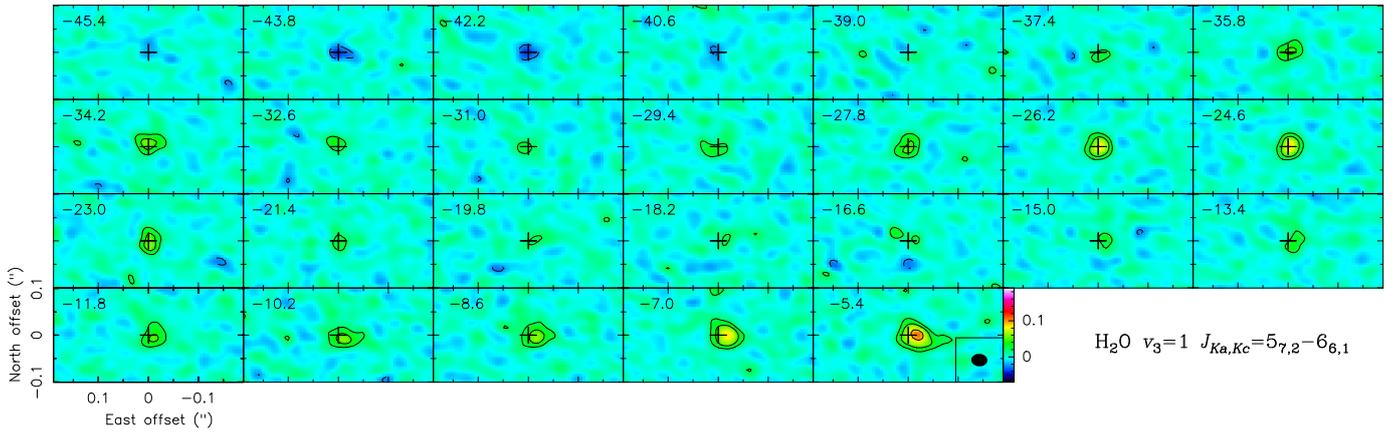}}}
  \caption{Same as \fig\ref{fig.map.b6_13co21} but for H$_2$O $v_3$=1 $J_{K_{a},K_{c}}$=5$_{7,2}$--6$_{6,1}$ using a first level of $\pm$22\mJyBeam. Absorption and emission of the SiO $v$=9 $J$=17--16 and $^{12}$CO $v$=0 $J$=6--5 lines are shown from -45.4 to -40.6\kms{} and from -11.8 to -5.4\kms{} LSR, respectively.}
    \label{fig.map.b9_H2Ov3_1_572_661}
\end{figure*}

\begin{figure*}[htpb]
  \centering{\resizebox{18.2cm}{!}{
  \includegraphics{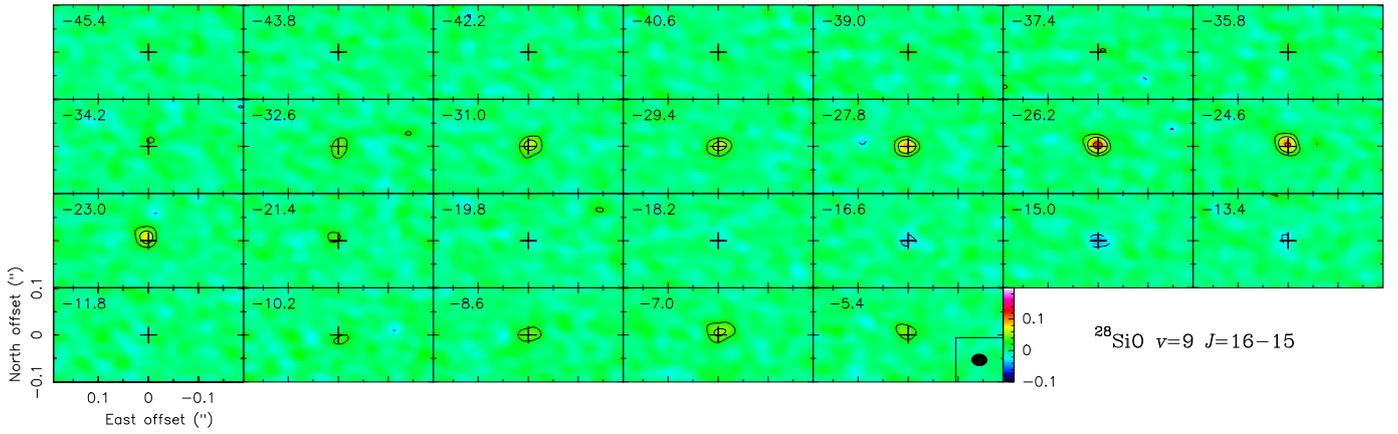}}}
  \caption{Same as \fig\ref{fig.map.b6_13co21} but for SiO $v$=9 $J$=17--16 using a first level of $\pm$25\mJyBeam. Emission of the H$_2$O $v_3$=1 $J_{K_{a},K_{c}}$=5$_{7,2}$--6$_{6,1}$ line is shown from -10.2 to -5.4\kms{} LSR.}
    \label{fig.map.b9_SiOv9_17_16}
\end{figure*}

\begin{figure*}[htpb]
  \centering{\resizebox{18.2cm}{!}{
  \includegraphics{figures/mapLines/b9_C13S.eps}}}
  \caption{Same as \fig\ref{fig.map.b6_13co21} but for $^{13}$CS $J$=15--14 using a first level of $\pm$25\mJyBeam. }
    \label{fig.map.b9_C13S}
\end{figure*}

\clearpage
\newpage

\section{Central velocity channels}

In this section we show the maps of the three central velocity channels, $-$26.2, $-$24.6 and $-$23\kms{} of several molecules to check the different shapes exhibited in the inner region of the emission. To understand the effect of the continuum emission, the maps of the line are shown before and after the continuum subtraction.

\begin{figure*}[htpb!]
  \centering{\resizebox{15.0cm}{!}{
  \includegraphics{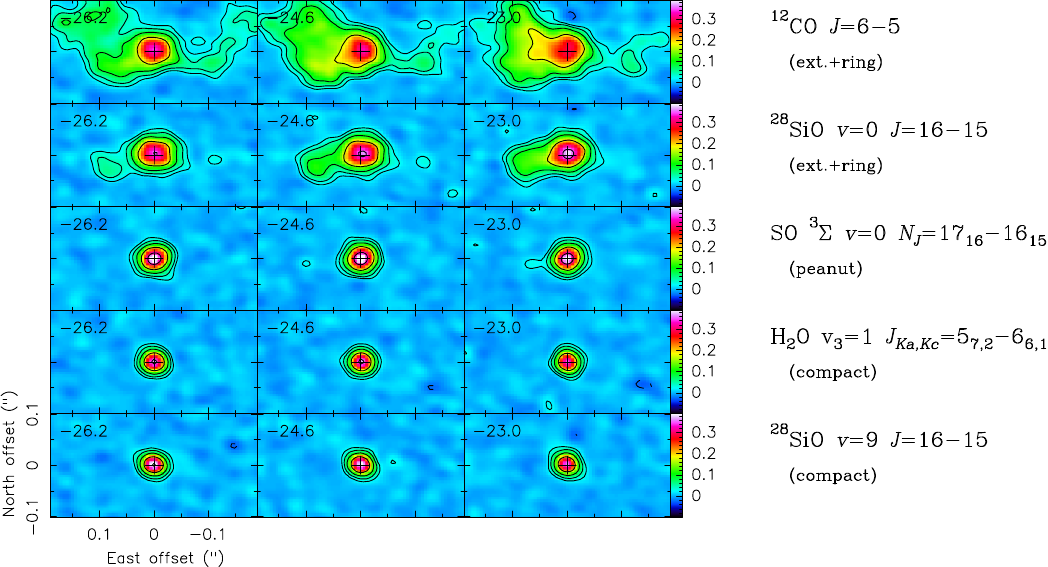}}}
  \caption{Central velocity channels of several molecules. The brightness units are \JyBeam. The continuum emission is not subtracted.}
    \label{fig.map.centralVelocitiesCont}
\end{figure*}

\begin{figure*}[htpb!]
  \centering{\resizebox{15.0cm}{!}{
  \includegraphics{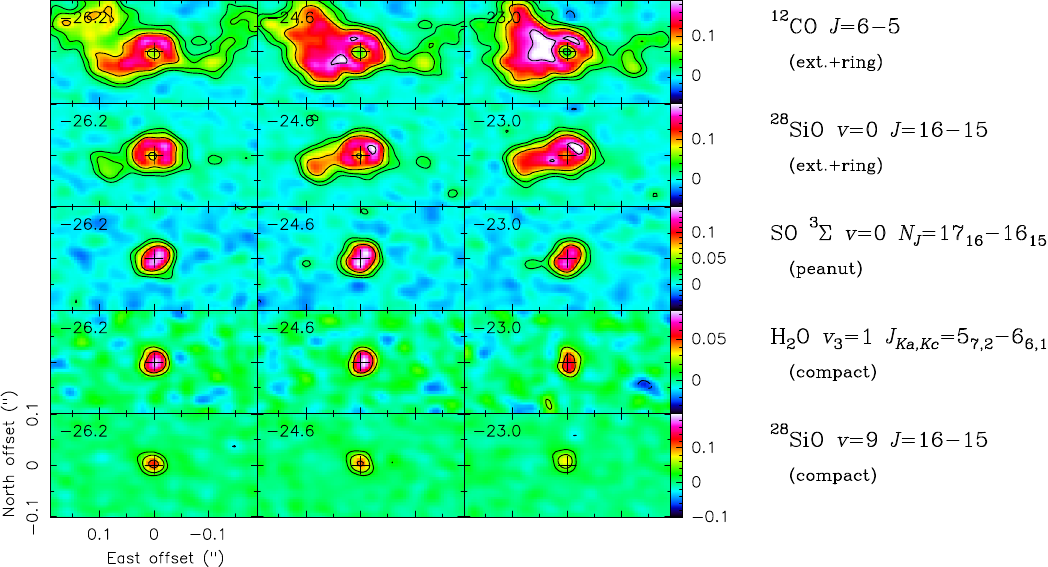}}}
  \caption{Same as \fig\ref{fig.map.centralVelocitiesCont} but with continuum emission subtracted. We call attention to the different brightness scales.} 
    \label{fig.map.centralVelocities}
\end{figure*}

\clearpage

\section{H30$\alpha$ and extended continuum emission map}\label{app.h30a}

\begin{figure}[htpb!]
   \centering{\resizebox{9cm}{!}{
   \includegraphics{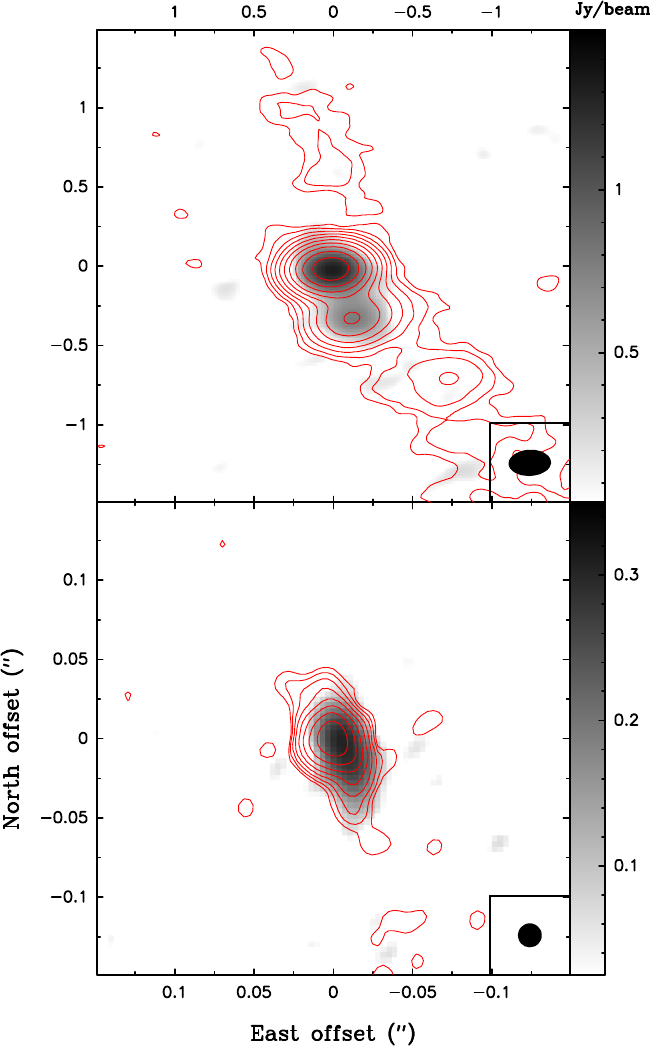}}}
   \caption{Low-resolution \textit{upper panel} and high-resolution (\textit{bottom panel}) maps of continuum (red contours) and H30$\alpha$ (gray scale) observed toward \raqr{}. For both low- and high-resolution maps, the contours are logarithmic, with a step of a factor of two and a first level of $\pm$0.07 and $\pm$0.1 \mJyBeam respectively. The HPBW is shown in the inset. We also note that the area shown in both panels is different.}
    \label{fig.map.conB6H30aLowHigh}%
    \end{figure}

\clearpage

\clearpage

\end{document}